\documentclass{emulateapj}

\usepackage{subfigure,epsfig,url,bm,amsmath,cases,color,soul}

\begin{document}

\title{More Realistic Planetesimal Masses Alter Kuiper Belt Formation Models and Add Stochasticity}

\author{Nathan A. Kaib}
\affiliation{Planetary Science Institute, 1700 E. Fort Lowell, Suite 106, Tucson, AZ 85719, USA}
\affiliation{HL Dodge Department of Physics \& Astronomy, University of Oklahoma, Norman, OK 73019, USA}
\author{Alex Parsells}
\affiliation{Department of Earth \& Environmental Sciences, Columbia University, New York, NY 10964, USA}
\affiliation{HL Dodge Department of Physics \& Astronomy, University of Oklahoma, Norman, OK 73019, USA}
\author{Simon Grimm}
\affiliation{University of Bern, Physikalisches Institut; Gesellschaftsstrasse 6, 3012 Bern, Switzerland}
\affiliation{ETH Z\"{u}rich, Institute for Particle Physics and Astrophysics, Wolfgang-Pauli-Strasse 27, 8093 Z\"{u}rich, Switzerland}
\author{Billy Quarles}
\affiliation{Department of Applied Mathematics and Physics, Astronomy, Geosciences \& Engineering Technology, Valdosta State University, Valdosta, GA 31698, USA}
\author{Matthew S. Clement}
\affiliation{Johns Hopkins APL, 11100 Johns Hopkins Road, Laurel, MD 20723, USA}

\begin{abstract}

Much of the modern Kuiper belt is thought to be the result of Neptune's migration through a primordial belt of planetesimals, possibly interrupted by an instability amongst the giant planets. While most prior work has employed massless test particles to study this Kuiper belt formation process, we perform simulations here that include the gravitational effects of the primordial planetesimal belt consisting of $\sim$10$^5$ massive bodies. In our simulations, Neptune unlocks from resonance with the other giant planets and begins to migrate outward due to interactions with planetesimals before a planetary orbital instability is triggered, and afterward, residual Neptunian migration completes the formation of the modern Kuiper belt. Compared to past simulations using massless test particles, our present work exhibits a number of notable differences. First, Neptune's planetary resonance unlocking requires the Neptunian 3:2 mean motion resonance to sweep much of the primordial disk interior to 30 au prior to the giant planet instability. This generates a pre-instability population of planetesimals that is significantly lower in semimajor axis, eccentricity, and inclination than predicted by many prior models, and this effect persists after the instability. Second, direct scattering between Pluto-mass bodies and other small bodies removes material from Neptunian resonances significantly more efficiently than the resonant dropout resulting from small changes in Neptune's semimajor axis during scattering between Pluto-mass bodies and Neptune. Consequently, the primordial population of Pluto-mass bodies may be as few as $\sim$200 objects. Finally, our simulation end states display a wide variety of orbital distributions, and straightforward relationships between final bulk Kuiper belt properties and Neptune's migration or initial planetesimal properties largely elude us. In particular, we find that the rapid, stochastic planetary orbital evolution occurring during the giant planet instability can significantly alter final Kuiper belt properties such as its inclination dispersion and the prominence of resonant populations. This complicates efforts to use modern Kuiper belt properties to confidently constrain early solar system events and conditions, including planetary orbital migration and the primordial Kuiper belt's characteristics. 

\end{abstract}


\section{Introduction}

Given its relative isolation from the planets, much of the modern Kuiper belt has only undergone modest dynamical evolution since its formation, and its properties and orbital architecture offer windows into the past conditions and events in the early solar system that accompanied this formation. As the Sun's giant planets scatter small planetesimals, the orbits of Saturn, Uranus, and Neptune migrate away from the Sun while Jupiter moves Sunward \citep{fernip84, hahnmal99}. Pluto's orbit as well as the abundance of objects trapped in Neptunian mean motion resonances (MMR) suggest that the modern Kuiper belt resulted from the dynamical dispersal of a much more massive primordial belt through which Neptune migrated early in the solar system's history \citep{mal93, mal95, chiang03}. Specifically, dynamical simulations have shown that a reservoir broadly analogous to our modern Kuiper belt is produced if Neptune migrates from $\sim$25 au to 30 au over the course of 100--300 Myrs through a disk of particles extending out to 30 au \citep[e.g.][]{nes15a, volkmal19}. The major exception that this origin process does not seem to explain is a subpopulation of the Kuiper belt known as the cold classical Kuiper belt, whose mostly circular, nearly coplanar orbits coupled with its unique physical properties (color, size distribution, and binary fraction) are taken as evidence that it instead formed in-situ \citep{tegrom00, noll08, fras14}. The cold classical belt's in-situ origin is further supported by New Horizons' observations of Arrokoth's physical properties, which are consistent with its continuous existence in a low-density, dynamically calm environment \citep{spenc20, mckin20}. 

Although it is generally agreed that Saturn's, Uranus', and Neptune's initial orbits were significantly closer to the Sun, the nature and timescale of their migration to their modern orbits is still unsettled. Generating Pluto's orbital eccentricity from a circular orbit requires an outward Neptunian migration of at least $\sim$5 au \citep{mal93}. However, this migration does not necessarily need to be smooth. If Saturn initially sits Sunward of Jupiter's 2:1 MMR, then the two gas giants must pass over their 2:1 MMR (and possibly other low-order resonances) during their divergent migration. This resonance crossing quickly excites the ice giants onto crossing orbits, touching off an episode of planet-planet scattering \citep{thom99, tsig05}. In such a scenario, Jupiter and Saturn's continued smooth migration past their 2:1 MMR will quickly overexcite the terrestrial planets' orbits and asteroid belt inclinations unless the gas giants' divergence is rapidly accelerated through direct scatterings with an ice giant \citep{bras09, kaibcham16, deien18, clem19}. In this case, gas giant-ice giant scattering events are very likely to result in the ejection of an ice giant \citep[implying the early solar system possessed 3+ ice giants;][]{nes11, nesmorb12} and could coincide with a sudden jump in Neptune's semimajor axis from ice giant-ice giant scattering \citep{nes15b}. 

Simulations modeling the formation of the Kuiper belt in concert with a dynamical instability among the giant planets typically simulate a primordial Kuiper belt consisting of tens of M$_{\oplus}$ of material with $\sim$10$^{3-4}$ particles because higher particle numbers result in prohibitively long run times on single processor cores \citep[e.g.][]{nesmorb12, clem18}. Because the trapping efficiency into the modern Kuiper belt is well below 1\% \citep{nes15a}, these particle numbers do not allow for a detailed direct comparison between the observed Kuiper belt and the final states of simulated systems. Thus, to generate single simulated Kuiper belts that are populous enough for statistical analysis, most recent works have forced the giant planets to follow predetermined orbital evolution sequences while the primordial belt is modeled with massless test particles that only respond to the gravity of the Sun and planets \citep[e.g.][]{mal95, lev08, nes15a, kaibshep16, volkmal19}. In this way, batches of test particles subjected to the same sets of perturbative forces can be split onto many different processing cores, with the results co-added to yield a statistically significant number of simulated Kuiper belt bodies. 

This simulation technique has replicated numerous features of the Kuiper belt. In particular, inclinations can be excited to of order $\sim$10$^{\circ}$ or higher if Neptune migrates $\sim$5 au or more to reach its modern orbit \citep{nes15a}. In addition, the ratio of Plutinos (objects trapped in Neptune's 3:2 MMR) to Hot belt objects (eccentric, inclined non-resonant objects between Neptune's 3:2 and 2:1 MMRs) can be lowered to the observed value if Neptune's migration is given a slight graininess that would result from scattering 1000--4000 Pluto-mass bodies during its migration \citep{nes16}. Moreover, although the cold classical belt likely formed in-situ, Neptune's 2:1 MMR must sweep across it during planet migration, stirring it. \citet{nes15b} showed that if this sweeping is suddenly interrupted by a jump in Neptune's semimajor axis coinciding with a planetary orbital instability, then an overdensity of unstirred cold belt material will be left untouched by Neptune's 2:1 MMR, which may explain the observed overdensity of cold belt objects near 44 au \citep{pet11}. Finally, if this sudden Neptunian jump also excited Neptune's eccentricity, then a small population of low-inclination ($i<10^{\circ}$), high-perihelion ($q>35$ au) Kuiper belt objects should be deposited at semimajor axes between 50 and 60 au, which Kuiper belt surveys have uncovered \citep{bann18, nes21}. 

Of particular relevance to the work we present here are test particle simulations performed in \citet{kaibshep16}. These simulations were largely a replication of model subsets presented in \citet{nes15a} and \citet{nes16} as Neptune migrated through a test particle belt from 24--30 au at different speeds, both smoothly and ``grainily.'' The focus of \citet{kaibshep16} was how Neptune's more distant resonances (5:2, 3:1, etc.) can leave behind a trail of escaped bodies stranded at high perihelion ($q\gtrsim40$ au) and high inclination ($i\gtrsim25^{\circ}$) as Neptune migrates \citep{nes16b, gom05}. These resonance trails are most prominent if Neptune took at least $\sim$100 Myrs to reach its modern orbit and had a level graininess consistent with $\sim$2000 primordial Pluto-mass bodies. While resonant trails are not the focus of our current work, the ``grainy slow'' run (a 300-Myr-long grainy migration sequence for Neptune) is used to compare our new simulations with past test particle runs. 

Although simulation work using test particles has greatly increased our understanding of the relationships between modern Kuiper belt structure and the early solar system, it is not without its limitations. The nature and speed of planetary migration depend on the population size and orbital distribution of Kuiper belt bodies at any given time, and this style of simulation does not allow for such feedback. In addition, modeling the formation of the Kuiper belt with massless test particles ignores the self-gravity of the Kuiper belt. Because the modern belt's predecessor possessed perhaps tens of Earth-masses of material, including its self-gravity may alter the results of simulations compared to those that do not include it \citep{nesmorb12, quarkaib19}. 

The advent of GPU-accelerated mixed variable symplectic integrators allows us to perform high-particle-number simulations of Kuiper belt formation that include the gravitational influence of the belt, including the ability of its primordial population of Pluto-mass bodies to dynamically stir the belt \citep{wishol91, cham99, grimmstad14, fanbat17, grimm22}. We present such work here. We start Jupiter, Saturn, and three ice giants in a favored resonant configuration \citep{nesmorb12} surrounded by a belt of $\sim$10$^5$ massive bodies and then allow the systems to evolve for 4 Gyrs, yielding an analog to our modern Kuiper belt in each case. This simulation technique is also not without drawbacks, as we have no control over the detailed evolution of the giant planets after the simulations are initialized, and some of them undoubtedly evolve in manners not relevant to our actual solar system. Nonetheless, in what follows, we find that  this technique offers us an opportunity to examine prior simulation results in a new manner and to characterize dynamical processes that were under-explored in previous Kuiper belt studies. 

The remainder of our paper is structured as follows: In Section \ref{sec:meth}, we describe the details of our simulations, including our numerical pipeline and initial conditions. In Section \ref{sec:res}, we discuss our simulation results. This begins with a discussion of the planets' orbital evolution in subsections \ref{sec:mig} and \ref{sec:plorbs}. Following this, we discuss the orbital evolution of the Kuiper belt in subsections \ref{sec:examp}--\ref{sec:FinKB}, giving special attention to the Kuiper belt's inclination distribution and population of Pluto-mass bodies in subsection \ref{sec:FinKB}. Finally, we summarize our conclusions in Section \ref{sec:con}. 

\section{Dynamical Simulation Methods}\label{sec:meth}

To perform our simulations, we use the GENGA N-body integrator \citep{grimmstad14}. This code utilizes the hundreds to thousands of CUDA cores within a single modern NVIDIA GPU to integrate single systems with much larger numbers of self-interacting bodies than most CPU-based integrators can perform within a reasonable timescale. In principle, the N-body integrator behaves comparably to the MERCURY hybrid integrator \citep{cham99}, symplectically integrating bodies within democratic heliocentric coordinates \citep{dun98} and smoothly switching over to a Bulirsch-Stoer integration \citep{stobu80} during close encounters ($<3$ Hill Radii) between massive bodies. 

Our simulations of Kuiper belt formation begin with five giant planets in a 3:2, 3:2, 2:1, 3:2 resonant chain (from inside to outside). In our chains, Jupiter resides closest to the Sun, followed by Saturn. While Jupiter and Saturn are set to their modern masses and radii, exterior to Saturn are three 15.8 M$_{\oplus}$ planets that are taken to approximate Uranus and Neptune. (These ice giant masses are the two real planets' average mass, as we do not know a priori which planet will be lost during an instability.) Instability simulations within our chosen resonant chain have been performed before \citep{nesmorb12, kaibcham16}, but in our simulations Neptune initially sits at 20.4 au instead of the 22.2 au used in prior works with this resonant chain. This is done because low-$N$ (1000-particle) runs found that Jupiter finished simulations too far from the Sun \citep{kaibcham16}. In the simulations presented in this work, Jupiter finishes with a median semimajor axis of 5.14 au, near the observed value of 5.2 au. Relative to many other tested 5-planet chains, our chosen resonant chain has higher likelihoods of generating other key features of the gas giants' orbital evolution \citep{nesmorb12}, although potentially more successful chains have been uncovered recently \citep{clem21a, clem21b}. 

In addition, a 20 M$_{\oplus}$ planetesimal disk is initialized between the final ice giant and 30 au. In half of our runs, the innermost planetesimal semimajor axis is set to 21.4 au (1 au beyond the outermost ice giant), and in the other half, it is set to 23.4 au. The planetesimals' initial orbital eccentricities are randomly sampled from a uniform distribution between 0 and 0.01, and the inclinations are randomly sampled from a uniform distribution between 0 and 1$^\circ$. As will be seen, planetary forcing and planetesimal self-stirring quickly excite inclinations above these values, and we wish to avoid early planetary instabilities or excited final Kuiper belts that are purely a consequence of our initial conditions. Meanwhile, planetesimal semimajor axes are randomly sampled from a uniform distribution truncated at 30 au, generating an $a^{-1}$ surface density profile. Prior studies of alternate surface densities exhibit little effect on instability outcomes \citep{batbrown10, nesmorb12}. Finally, mean anomalies, longitudes of ascending node, and arguments of perihelion are all randomly sampled from uniform distributions between 0 and 360$^{\circ}$. 

Each of our planetesimal disks contains between 0 and 2500 Pluto-mass bodies. These bodies' gravitational interactions with themselves and every other particle are fully computed at each simulation timestep. We refer to such bodies as ``fully active.'' The remainder of the disk mass is comprised of $\sim$0.09 Pluto-mass bodies. These sub-Plutos are ``semi-active'' bodies that do not interact with one another, but they do exert gravitational influence on all other more massive bodies (and feel those bodies' gravity). Depending on the number of Pluto-mass bodies, each planetesimal disk contains between 75,000 and 100,000 sub-Plutos (which comprise 75--100\% of our disks' masses). Table \ref{tab:IC} provides an overview of the initial conditions of each of our 12 simulations. 

\begin{table}
\begin{tabular}{c c c c}
\textbf{Simulation Initial Conditions} & & & \\
\end{tabular}
\centering
\begin{tabular}{c c c c c c}
\hline
Sim & $m_{disk}$ & $N_{Pluto}$ & $N_{subPluto}$ & $a_{min}$ & $a_{max}$ \\
 & ($M_{\oplus}$) & & & (au) & (au) \\
\hline
0Pa & 20 & 0 & 100000 & 23.4 & 30 \\
0Pb & 20 & 0 & 100000 & 23.4 & 30 \\
200Pa & 20 & 200 & 97800 & 23.4 & 30 \\
200Pb & 20 & 200 & 97800 & 23.4 & 30 \\
400Pa & 20 & 400 & 95600 & 21.4 & 30 \\
400Pb & 20 & 400 & 95600 & 21.4 & 30 \\
700Pa & 20 & 700 & 93000 & 21.4 & 30 \\
700Pb & 20 & 700 & 93000 & 21.4 & 30 \\
1000Pa & 20 & 1000 & 89000 & 23.4 & 30 \\
1000Pb & 20 & 1000 & 89000 & 23.4 & 30 \\
2500Pa & 20 & 2500 & 75000 & 21.4 & 30 \\
2500Pb & 20 & 2500 & 75000 & 21.4 & 30 \\
\hline
\end{tabular}
\caption{Table of simulation initial conditions. From left to right, the columns are as follows: (1) simulation name, (2) total mass of primordial belt, (3) initial number of Pluto-mass bodies, (4) initial number of sub-Pluto bodies, (5) minimum semimajor axis of primordial belt bodies, (6) maximum semimajor axis of primordial belt bodies.}
\label{tab:IC}
\end{table}

Our systems are integrated for 4 Gyrs with a timestep of 200 days. Particles are removed from the simulation if their heliocentric distance exceeds 1000 au or upon collision with fully active particles or the Sun (whose radius is inflated to 0.5 au). Interactions between the planetesimals and the planets unlock the planets from their resonant chain and drive a migration of their semimajor axes. This eventually triggers an orbital instability amongst the planets. There is no guarantee that such an instability will result in the system collapsing down to an outer planetary configuration that resembles our own, and past works have run many sets of such simulations to generate a small ensemble of ``solar system-like'' architectures \citep[e.g.][]{nesmorb12, kaibcham16, clem18}. In this work, we do not have this option because the simulation sizes and runtimes ($\sim3$ months across a variety of modern NVIDIA GPUs) make this infeasible. Instead, when a system undergoes an instability, the simulation is stopped and different realizations are restarted at the last time output prior to the instability's onset. These realizations are all identical, except that the Cartesian positions and velocities of innermost ice giant are randomly shifted by 1 part in 10$^9$ in each realization. This shift is significantly smaller than the energy error of the integrator \citep{cham99}. Because the sequence of planetary encounters during the orbital instability is highly chaotic, these small changes in initial conditions generate wildly different post-instability states. These realizations' post-instability states are generated until we arrive at two different ``solar system-like'' outcomes wherein Jupiter and Saturn have crossed their 2:1 MMR, and the system contains four total planets. Only these post-instability states are integrated for the full 4 Gyrs and comprise our simulation set. We find that it typically requires of order 10 instability realizations to yield a ``solar system-like'' outcome. In our simulation naming scheme, our two solar system-like outcomes are designated names ending in `a' and `b,' but the assignment of each particular letter carries no significance. This instability-cloning procedure saves a significant amount of computing time. For instance, our 700Pa simulation required 17 days of computing to reach the point of instability, and repeating this $\sim$10 times to yield a solar system-like instability outcome would require nearly 0.5 years of GPU computing time.

Examining the final outcomes of our 4-Gyr simulations, we immediately noticed that Uranus and Neptune migrated across their 2:1 MMR in 11 of our 12 systems. This resonance-crossing event can dramatically reshape the orbital architecture of the Kuiper belt in our simulated systems \citep{grahamvolk24}. Moreover, in our actual solar system, the two planets' period ratio is $\sim$1.96, and there is no speculation that this resonance-crossing has occurred. Thus, to prevent our simulated Kuiper belts from being reshaped by this event, we repeat 11 of our successful post-instability realizations with Uranus' immediate post-instability semimajor axis shifted away from the Sun by an additional $\sim$0.5--1.5 au to prevent a 2:1 crossing in our realizations. Simulation 2500Pb (see Table \ref{tab:IC}) is the only system whose post-instability realization did not have to be repeated with an augmented Uranian orbit. 

Our manual movement of Uranus' orbit is obviously not a pre-planned numerical procedure and results from our inability to predict Uranus' orbital evolution prior to running the actual simulations. To gauge how this orbit resetting alters the bulk evolution of our systems, we measure the median inclination and eccentricity of the planetesimals in each system 10 Myrs after Uranus' orbit is reset and compare these two values against the medians seen in each unaltered counterpart system. This amount of time is short enough that our unaltered systems have not yet crossed the Uranus-Neptune 2:1 MMR, but also long enough for our altered systems to feel the dynamical effects of their Uranus orbit resetting. In our 11 systems with an altered Uranus, we find a mean difference in the median inclination of our altered and unaltered systems of 0.43$^{\circ}$ (with a maximum difference of 1.27$^{\circ}$), and a mean difference in the median eccentricity of our altered and unaltered systems of 0.014 (with a maximum of 0.05). As we will see in Section \ref{sec:FinKB}, there is a substantially larger variation in median inclination ($\pm$2$^{\circ}$) and median eccentricity ($\pm$0.1) across our whole ensemble of systems as they emerge from their planetary instabilities (see Figure \ref{fig:selfstir3}). Thus, while the manual movement of Uranus does have a modest effect on our systems' dynamical evolution (as we should expect), it is not the main driver of the range of results we will discuss in the proceeding sections.

\section{Results and Discussion}\label{sec:res}

\subsection{Neptune Migration Rate}\label{sec:mig}

Many past works modeling Kuiper belt formation force Neptune's semimajor axis migration with a fictitious drag force that declines exponentially as Neptune's modern semimajor axis is approached \citep[e.g.][]{mal95, nes15a, kaib19}. This prescribed force is implemented because Kuiper belt objects are treated as massless test particles in order to attain high enough particle numbers to resolve the final belt's structure. Meanwhile, our GPU-accelerated simulations presented here contain $\sim$$10^5$ bodies, and actual gravitational scattering between Neptune and primordial Kuiper belt objects drives Neptune's outward migration. 

\begin{figure*}
\centering
\includegraphics[scale=0.65]{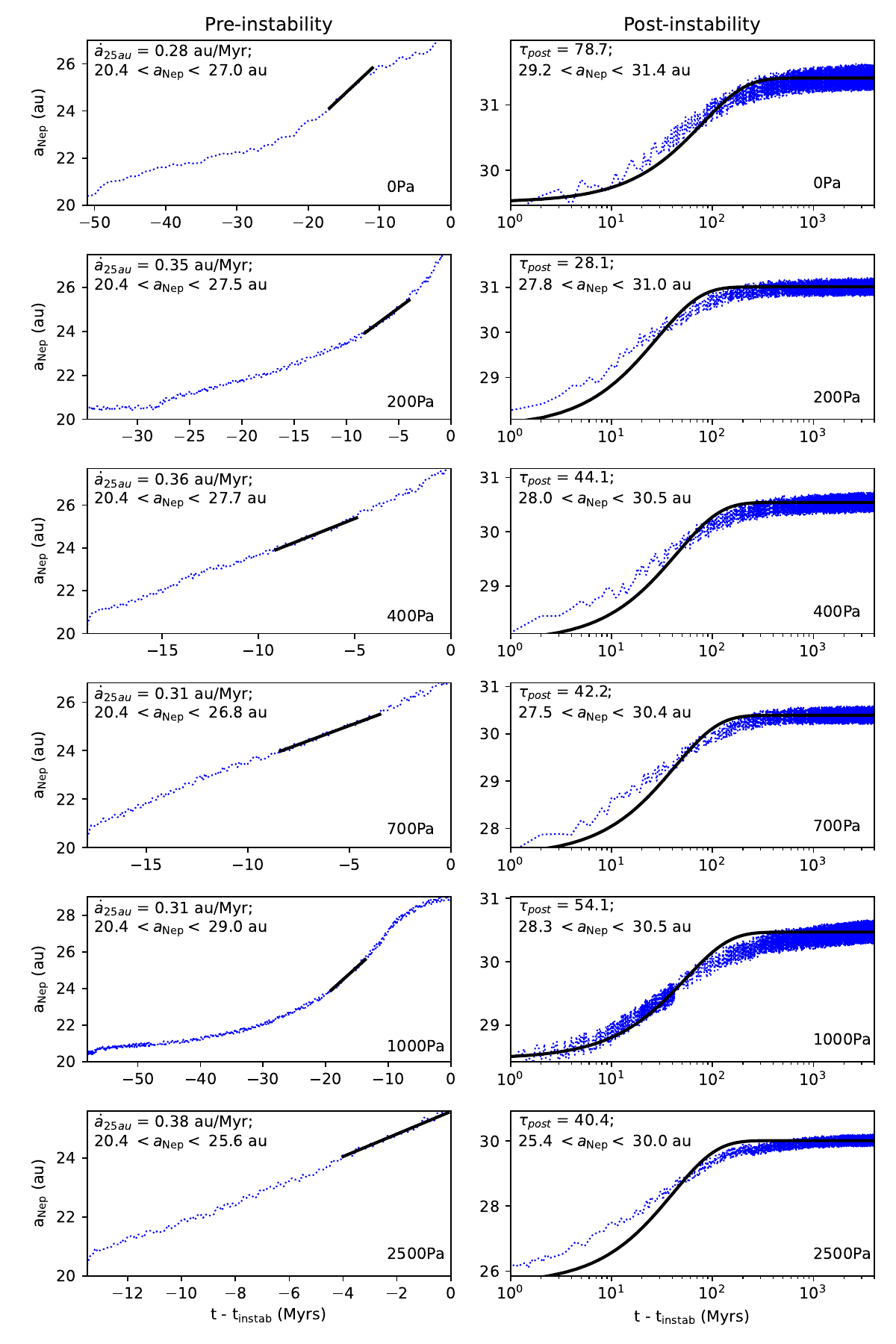}
\caption{{\bf Left column:} The evolution of Neptune's semimajor axis vs time before the giant planet instability ({\it dotted blue}) as well as the best-fit linear profile for the simulated evolution between 24--25.6 au ({\it solid black}). {\bf Right column:} The evolution of Neptune's semimajor axis vs time after the giant planet instability ({\it dotted blue}) as well as the best-fit exponential decay profile for the simulated evolution ({\it solid black}). Individual simulations are specified with plot labels. The pre- and post-instability migration range is noted in plot annotations as well as each migration's best-fit profile.}
\label{fig:migfits}
\end{figure*}

In Figure \ref{fig:migfits}, we compare Neptune's semimajor axis evolution seen in our simulations with a forced exponentially decaying migration rate. To keep the number of panels reasonable, this figure only includes the `a' post-instability simulations (0Pa, 200Pa, etc.) and not the `b' ones, but the behavior is largely the same in the `b' simulations. Since recent past works employ an early migration rate and a different, late migration rate \citep[e.g.][]{nes15b, kaibshep16}, we make the semimajor axis comparisons in two separate epochs: before each system's giant planet instability and after each system's giant planet instability. The left column of Figure \ref{fig:migfits}, shows Neptune's migration from 20.4 au until the giant planets' orbits become unstable. Here we see that the location of the disk's inner edge determines the first phase of behavior. In cases where the inner edge is at 21.4 au (400Pa/b, 700Pa/b, and 2500Pa/b), the migration begins immediately, and the rate of change is nearly constant with time. In the other cases where the inner disk edge is at 23.4 au (0Pa/b, 200Pa/b, and 1000Pa/b), the migration slowly accelerates until the instability occurs. 

In these pre-instability simulations, we initially attempted to fit semimajor axis evolution with an exponential profile, since numerous recent Kuiper belt formation models prescribe such a profile to Neptune's pre-instability migration \citep{nes15a, kaib19}. However, the left column of Figure \ref{fig:migfits} clearly shows that the pre-instability migration seen in our simulations does not resemble an exponential. To gauge the speed of pre-instability migration, we instead use a linear fit when Neptune is between 24 au \citep[Neptune's starting semimajor axis used in many recent works;][]{kaibshep16} and 25.6 au (the closest semimajor axis at which any of our simulated migrations are interrupted by an instability). Doing this we see that the 0Pa/b simulations exhibit the slowest pre-instability migration rates (0.28 au/Myr) and the 2500Pa/b exhibit the fastest rates ($\sim$0.37 au/Myr), but they are all within $\sim$30\% of one another and center around values near 0.3 au/Myr. It is tempting to explain the slightly slower migration rates in 0Pa/b with their lack of Pluto-mass bodies, which can drive more planetesimals into Neptune encounters via self-excitation of the disk. However, the slower migration in 0Pa/b is modest (see 700Pa/b and 1000 Pa/b), and it may simply be statistical variance among our small set of simulations.

We also study the behavior of Neptune's migration after the giant planet instability. This behavior is shown in the righthand column of Figure \ref{fig:migfits}. Fitting this phase of semimajor axis evolution with an exponential, we see that the timescale of migration lengthens significantly after the instability, consistent with past works \citep{nes15b}. This second phase of migration is fitted with decay constants ranging near $\sim$28--44 Myrs in most of our simulations. The exceptions to this are the 0Pa/b simulations, which have timescales of $\sim$50 and $\sim$80 Myrs, and the 1000Pa/b simulations, which have timescales of $\sim$54 and $\sim$65 Myrs. In the case of 0Pa/b, the lack of disk-self stirring may explain the slightly slower migration timescales. Meanwhile, the 1000Pa/b simulations experience the longest delay before the giant planet instability takes place. In these simulations, it occurs after 58 Myrs when Neptune has reached 29 au, which is very near the original outer edge of the primordial belt at 30 au. Thus, after the 1000 Pa/b instability, there are very few bodies left to induce residual migration.

Once again, Figure \ref{fig:migfits} shows that the post-instability migration of Neptune is not particularly well-fit by an exponential. In the post-instability regime, the exponential profile evolves too slowly immediately after the instability and then too quickly in later epochs. Such behavior is expected from prior simulations of giant planet instabilities using much lower particle numbers \citep{nesmorb12, nes15b}. We can also see this effect when we measure the amount of time it takes for Neptune to migrate within 0.1 au of its final median semimajor axis measured between $t=$ 3.9--4.0 Gyrs. In the case of a prescribed, forced migration used in the grainy slow run of \citet{kaibshep16}, Neptune reaches this semimajor axis within 230 Myrs. The timescale values for our current set of simulations are shown in Table \ref{tab:nepmig}. Only one of our current simulations (400Pb) features Neptune approaching its final semimajor axis this quickly ($t=190$ Myrs). All of the other simulations feature a more drawn out residual migration of Neptune, and Neptune comes within 0.1 au of its final semimajor axis with a median time of 340 Myrs. This median is $\sim$100 Myrs longer than the \citet{kaibshep16} grainy slow run, which is among the slowest migration scenarios previously considered for Kuiper belt formation \citep{nes15a}. 

Figure \ref{fig:migfits} also shows that Neptune's post-instability migration often never fully stops and minor semimajor axis evolution is visibly apparent until $\sim$1 Gyr. In addition, our simulations exhibit a spectrum of different final Neptunian semimajor axes between 29.5 and 31.4 au (see Table \ref{tab:nepmig}). Thus, the outer edge of our disks only sets the stopping position of Neptune within $\pm$1 au, and the exact stopping position is likely related to the properties of the residual disk after Neptune completes most of its migration \citep{gom04}. 

\begin{table*}
\centering
\begin{tabular}{c}
\textbf{Neptune Migration}\\
\end{tabular}
\\
\centering
\begin{tabular}{c c c c c c c c c c}
\hline
Sim & $t_{instab}$ & $a_{instab}$ & $\Delta a_{N}$ & $e_{Nmax}$ & $i_{Nmax}$ & $a_{Nfinal}$ & $t_{stop}$ & $\dot{a}_{25au}$ & $\tau_{post}$ \\
 & (Myrs) & (au) & (au) & & ($^{\circ}$) & (au) & (Myrs) & (au/Myrs) & (Myrs) \\
\hline
0Pa & 51.0 & 27.0 & 2.2 & 0.12 & 1.62 & 31.4 & 570 & 0.28 & 78.7\\
0Pb & 53.0 & 26.8 & 0.0 & 0.02 & 1.08 & 29.6 & 630 & 0.28 & 48.2\\
200Pa & 34.8 & 27.5 & 0.3 & 0.03 & 0.87 & 31.0 & 270 & 0.35 & 28.1\\
200Pb & 34.7 & 27.7 & 0.0 & 0.03 & 0.34 & 30.6 & 360 & 0.35 & 31.9\\
400Pa & 18.9 & 27.7 & 0.3 & 0.05 & 0.81 & 30.5 & 440 & 0.36 & 44.1\\
400Pb & 18.7 & 27.5 & 0.0 & 0.06 & 0.51 & 29.5 & 190 & 0.36 & 31.1\\
700Pa & 17.7 & 26.8 & 0.3 & 0.03 & 0.22 & 30.4 & 310 & 0.31 & 42.2\\
700Pb & 19.2 & 27.1 & 0.4 & 0.11 & 1.00 & 29.8 & 320 & 0.31 & 38.3\\
1000Pa & 58.1 & 29.0 & -0.7 & 0.10 & 2.05 & 30.5 & 450 & 0.31 & 54.1\\
1000Pb & 58.1 & 29.0 & 0.0 & 0.02 & 0.63 & 31.4 & 300 & 0.31 & 64.5\\
2500Pa & 13.5 & 25.6 & -0.2 & 0.08 & 0.60 & 30.0 & 730 & 0.38 & 40.4\\
2500Pb & 13.6 & 25.6 & -0.6 & 0.08 & 1.06 & 29.5 & 320 & 0.36 & 29.1\\
\hline
\end{tabular}
\caption{Table of Neptune's orbital evolution. From left to right, the columns are as follows: (1) simulation name, (2) time at which the giant planet orbital instability occurs, (3) Neptune's semimajor axis when the giant planet orbital instability occurs, (4) the change in Neptune's semimajor axis during the giant planet orbital instability, (5) the maximum orbital eccentricity measured for Neptune, (6) the maximum orbital inclination measured for Neptune, (7) Neptune's final semimajor axis, (8) the time at which Neptune's semimajor axis reaches within 0.1 au of its final value, (9) Neptune's pre-instability semimajor axis migration rate from 24--25.6 au, and (10) the e-folding timescale of an exponential fit to Neptune's post-instability semimajor axis evolution.}
\label{tab:nepmig}
\end{table*}

\subsection{Final Orbits of the Giant Planets}\label{sec:plorbs}

The dispersal of a primordial planetesimal disk drives orbital evolution among the rest of the giant planets as well. In Table \ref{tab:plorbs}, we list key parameters in the final architectures of the giant planets after 4 Gyrs of evolution. In the first column we list the ratio of Saturn's to Jupiter's orbital period. In the real solar system, this ratio is $\sim$2.48, but our final simulated systems have a range of ratios spanning from 2.1--2.7, with a median value of 2.39. Although the locations of various secular resonances depend on this exact ratio, this has dynamical consequences primarily for the inner solar system rather than the Kuiper belt \citep[e.g.][]{bras09, walshmorb11, clem20}. 

However, the Kuiper belt's architecture is dependent upon the ratio of Neptune's to Uranus' orbital period, and we list these final ratios for our systems in the second column of Table \ref{tab:plorbs}. In our original simulations, all but one system (2500Pb) evolved above a period ratio of 2. Values in these original systems ranged from 1.97--2.31, with a median of 2.13. Consulting a large number of $\sim$1000-particle simulations from prior works \citep{clem18, clem21a, clem21b}, we find roughly half of all 4-planet instability outcomes finish with a Neptune-to-Uranus period ratio of under 2, so it is not clear whether this is an unrealized systematic issue with our particular 5-planet resonant chain or whether it is simply an issue of bad luck among our 12 simulated systems. Nevertheless, the crossing of Uranus and Neptune's 2:1 MMR can destabilize much of the resonant Kuiper belt population, limiting the utility of our simulation results, and is therefore a poor feature for simulations meant to study Kuiper belt formation \citep{grahamvolk24}.

Consequently, for every simulation except 2500Pb, we restart the system shortly after the giant planet instability, and Uranus' semimajor axis is shifted outward (while holding other orbital elements constant) to prevent a future crossing of its 2:1 MMR with Neptune. The rationale behind this was that while we cannot control the details of the giant planet instability, instability sequences do occur (as evidenced by 2500 Pb) that place Uranus on a post-instability orbit that prevents a 2:1 MMR crossing, and the final structure of the Kuiper belt is likely not strongly sensitive to a modest manual shift in Uranus' semimajor axis early in the solar system's history. These augmented systems are what are shown in Table \ref{tab:plorbs}. It is still difficult to predict the future evolution of systems after we manually shift Uranus' orbit, and we see that only a few of our systems attain ratios near the solar system's 1.96 value. Instead, they range from 1.83--1.98, with a median of 1.9. Nonetheless, these augmented systems stay on the correct side of the Uranian-Neptunian 2:1 MMR, and therefore provide better comparisons with the actual solar system. 

Table \ref{tab:plorbs} also includes measurements of the eccentricity of each giant planet. In the case of Jupiter and Saturn, we list $e_{55}$ and $e_{66}$, which are effectively the orbital eccentricities that Jupiter and Saturn would respectively possess in the absence of other planetary perturbations, as formulated within Laplace-Lagrange theory. As can be seen, Saturn's eccentricity ($e_{66}$) is higher than the solar system's value in all but two systems, and the median value of our simulated systems is 0.088, or nearly twice the solar system value. Meanwhile, Jupiter suffers the opposite problem, as its eccentricity is under-excited. Our simulations have a median $e_{55}$ value of 0.024, or barely half the solar system value, and no system exceeds the solar system value. This under-excitement of $e_{55}$ and over-excitement of $e_{66}$ is a known issue for simulations that generate the modern outer solar system via an orbital instability, and a clear solution beyond the invocation of low-probability events does not currently exist \citep{nesmorb12, clem21a, clem21b, clem21c}. Thus, our simulations are consistent with prior simulations of the giant planet orbital instability utilizing different numerical algorithms. Moreover, although our final gas giant orbital architectures are not identical to the actual solar system, they are comparable to others in the literature that have been regarded as solar system analogs and used to study the evolution of the terrestrial planets, the asteroid belt, and the Kuiper belt \citep[e.g.][]{lev08, deien18, nes21b}.

Uranus and Neptune also appear to have under-excited orbital eccentricities. The median value for Uranus' mean eccentricity is 0.022, and the median value for Neptune's mean eccentricity 0.005. These values are both half the values of the real solar system. However, in the case of the ice giants, the spectrum of results in our simulated systems brackets the solar system more nicely, as 5 of our 12 systems have Uranian and Neptunian mean eccentricities that exceed the observed solar system. The eccentricities of Uranus and Neptune are highly correlated, and these are the same 5 overexcited systems for each ice giant. 

Finally, Table \ref{tab:plorbs} also includes the time-averaged inclinations of each giant planet. Although the planetary inclinations are less intensely studied in the context of a giant planet instability, we do see some systematic differences between our solar system and our simulations. First, we note that the most highly inclined planet in each of our simulated systems is Saturn, whereas it is Uranus in the actual solar system. In addition, the ice giant inclinations are significantly lower than the gas giant inclinations in our simulations. This is not seen in the actual solar system, where Uranus has the highest inclination and Neptune's is still greater than Jupiter's inclination. Compared to the solar system, our ice giant inclinations are consistently significantly lower. Our median simulated value for Uranus is $\sim$0.2$^{\circ}$ (or 20\% of the solar system's value), and our median simulated value for Neptune is 0.075$^{\circ}$ (or $\sim$10\% of the solar system's value). Moreover, none of our simulated ice giant inclinations match or exceed the real ice giant values. It is not clear why there is this systematic discrepancy, but it warrants further study.

\begin{table*}
\centering
\begin{tabular}{c}
\textbf{Final Giant Planet Orbits}\\
\end{tabular}
\\
\begin{tabular}{c c c c c c c c c c c}
\hline
Sim & $\frac{P_{\rm S}}{P_{\rm J}}$ & $\frac{P_{\rm N}}{P_{\rm U}}$ & $e_{55}$ & $e_{66}$ & $\bar{e}_{\rm U}$ & $\bar{e}_{\rm N}$ & $\bar{i}_{\rm J}$ & $\bar{i}_{\rm S}$ & $\bar{i}_{\rm U}$ & $\bar{i}_{\rm N}$\\
 & & & & & & & ($^{\circ}$) & ($^{\circ}$) & ($^{\circ}$) & ($^{\circ}$) \\
\hline
0Pa & 2.70 & 1.88 & 0.016 & 0.112 & 0.105 & 0.017 & 1.54 & 3.76 & 0.53 & 0.11 \\
0Pb & 2.10 & 1.90 & 0.011 & 0.066 & 0.004 & 0.004 & 0.65 & 1.72 & 0.04 & 0.03 \\
200Pa & 2.53 & 1.88 & 0.042 & 0.070 & 0.085 & 0.015 & 0.41 & 1.01 & 0.80 & 0.21 \\
200Pb & 2.36 & 1.83 & 0.023 & 0.097 & 0.048 & 0.012 & 0.88 & 2.23 & 0.92 & 0.22 \\
400Pa & 2.26 & 1.95 & 0.025 & 0.070 & 0.007 & 0.004 & 0.51 & 1.32 & 0.06 & 0.02\\
400Pb & 2.42 & 1.85 & 0.042 & 0.096 & 0.087 & 0.015 & 1.16 & 2.93 & 0.27 & 0.21\\
700Pa & 2.20 & 1.98 & 0.021 & 0.045 & 0.008 & 0.004 & 0.16 & 0.40 & 0.15 & 0.03\\
700Pb & 2.46 & 1.87 & 0.041 & 0.103 & 0.024 & 0.005 & 0.56 & 1.42 & 0.14 & 0.04\\
1000Pa & 2.19 & 1.90 & 0.029 & 0.041 & 0.100 & 0.017 & 0.30 & 0.77 & 0.25 & 0.31\\
1000Pb & 2.43 & 1.97 & 0.040 & 0.102 & 0.019 & 0.004 & 0.86 & 2.16 & 0.86 & 0.13\\
2500Pa & 2.24 & 1.94 & 0.015 & 0.110 & 0.005 & 0.004 & 0.38 & 0.98 & 0.16 & 0.03\\
2500Pb & 2.62 & 1.97 & 0.003 & 0.080 & 0.008 & 0.005 & 0.29 & 0.73 & 0.05 & 0.04\\
\hline
\textbf{Solar System} & \textbf{2.48} & \textbf{1.96} & \textbf{0.044} & \textbf{0.048} & \textbf{0.044} & \textbf{0.010} & \textbf{0.37} & \textbf{0.90} & \textbf{1.02} & \textbf{0.67}\\
\hline
\end{tabular}
\caption{Table describing the final giant planet orbital architectures. From left to right, the columns are as follows: (1) simulation name, (2) ratio of Saturn's orbital period to Jupiter's orbital period, (3) ratio of Neptune's orbital period to Uranus' orbital period, (4) the amplitude of Jupiter's eccentricity eigenfrequency within Lagrange-Laplace theory, (5) the amplitude of Saturn's eccentricity eigenfrequency, (6) Uranus' mean orbital eccentricity, (7) Neptune's mean orbital eccentricity, (8) Jupiter's mean inclination, (9) Saturn's mean inclination, (10) Uranus' mean inclination, and (11) Neptune's mean inclination.  Values of these quantities for the real solar system are listed in the final table row.}
\label{tab:plorbs}
\end{table*}

\subsection{Illustrative Example of Kuiper Belt Formation}\label{sec:examp}

\begin{figure*}
\centering
\includegraphics[scale=0.6]{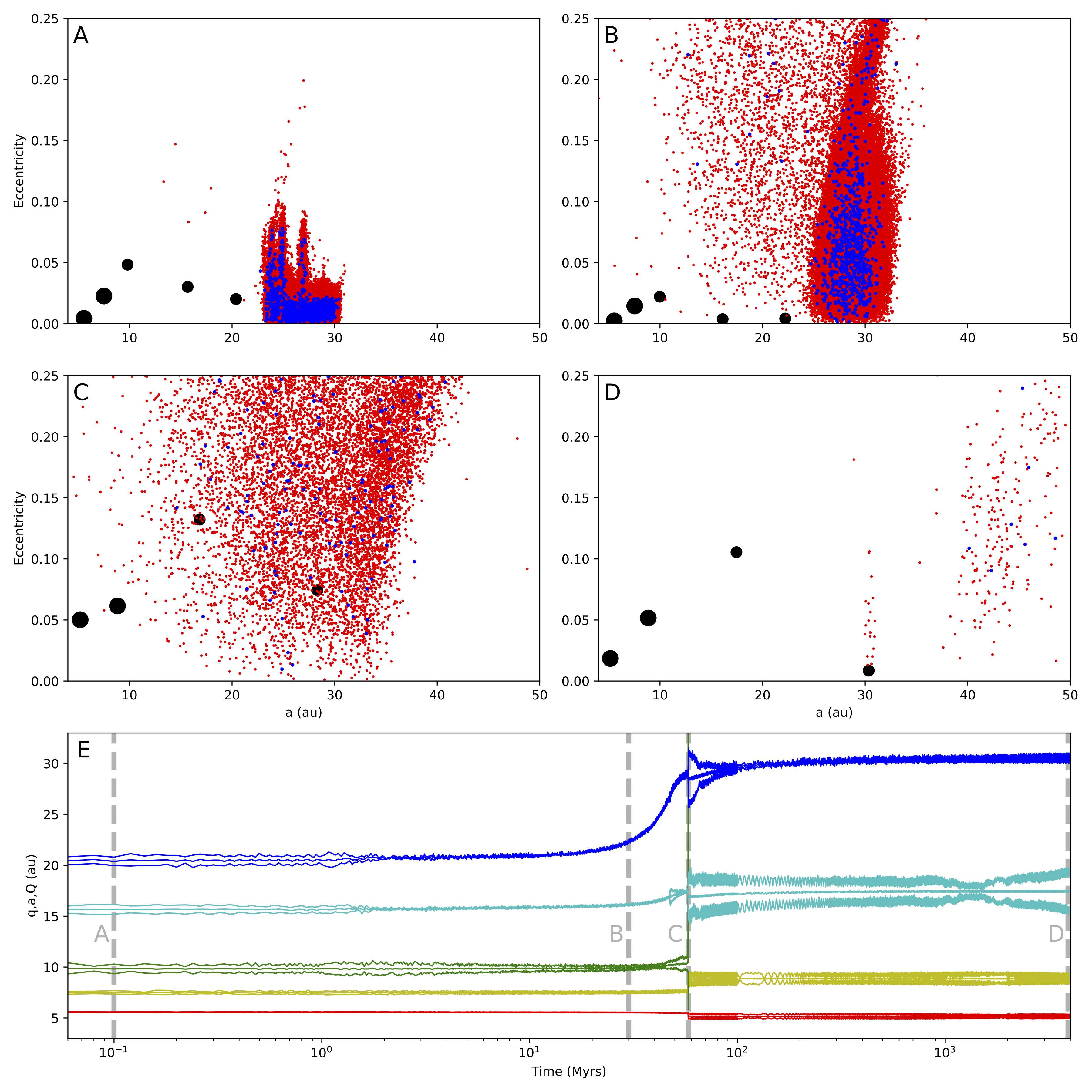}
\caption{Evolution of our 1000Pa system. {\bf A--D:} Eccentricities vs semimajor axes of all bodies at $t=$ 0.1 Myrs (panel A), 30 Myrs (panel B), 58.1 Myrs (panel C), and 3.9 Gyrs (panel D). Large black points mark Jupiter and Saturn, while smaller black points mark ice giants. Blue data points mark Pluto-mass bodies, and red points mark sub-Plutos. {\bf E:} Perihelion, semimajor axis, and aphelion of each planetary body is plotted vs time. Vertical dashed lines mark the times of each of the upper four panels. }
\label{fig:ExampleSim}
\end{figure*}

Figure \ref{fig:ExampleSim} shows the time evolution of our 1000Pa system, whose planetesimal belt initially contained 1000 Pluto-mass bodies and 90000 sub-Pluto bodies. Here we see that after just $10^5$ years Neptune excites the eccentricities of primordial belt bodies near its 3:2, 4:3, and 5:4 MMRs. In addition, we can see that outside of these resonances, scattering events with the Pluto-mass bodies enhance the eccentricities of many planetesimals well beyond their initial maximum value of 0.01. However, the dynamical friction from the sub-Plutos leads to less excitation among the Pluto-mass bodies. 

By 30 Myrs of evolution, Neptune has unlocked from resonance with the other giant planets and has migrated all the way to a semimajor axis near 22.5 au, or 1 au inside of the initial inner edge of the primordial planetesimal belt (23.4 au). At this point, most of the primordial belt remains on low eccentricity orbits ($e\lesssim0.2$) between 26 and 31 au. By comparing panels A and B, we see that the typical eccentricity of the planetesimal belt has grown by a factor of several. This growth has two main sources. The first is that the Pluto-mass bodies continue to scatter off of one another as well as the sub-Plutos. This leads to semimajor axis diffusion as well as eccentricity and inclination excitation of the planetesimal disk. In addition, as Neptune moves outward from 20.4 to 22.5 au, its exterior MMRs move over larger distances. The 3:2 MMR shifts from 26.7 au to 29.5 au, and the 5:4 MMR shifts from 23.7 au to 26.1 au (near the original position of the 3:2). In this manner, the entire primordial disk has been traversed by at least one first order MMR. 

It is also obvious from Panel B that the belt's eccentricity excitation generates a sizable population of planetesimals on planet-crossing orbits. This drives further migration of the giant planets, which eventually triggers a global instability amongst the planets at $t=58$ Myrs. This leads to the ejection of the innermost ice giant. Before its ejection, it excites Neptune's eccentricity to $\sim$0.1 and causes its semimajor axis to jump 0.5 au closer to the Sun (see Figure \ref{fig:ExampleSim}E). This also dramatically excites the planetesimal disk, which leads to its steady dispersal. This steady dispersal drives a residual migration of the giant planets with Neptune finally coming to rest near 30.5 au. (One should note that the evolution depicted in Figures \ref{fig:ExampleSim}D and  \ref{fig:ExampleSim}E is taken from the unmodified simulation version in which Uranus was not reset on a new post-instability orbital semimajor axis to prevent a 2:1 crossing with Neptune.) 

Figure \ref{fig:ExampleSim}D also shows the near-final state of the 1000Pa system. Besides an overexcited Uranus and an underexcited Jupiter, the planetary system looks broadly similar to the real outer planets. In addition, there is a reservoir that appears qualitatively similar to our modern Kuiper belt; Neptune possesses a significant population of Trojans, and the bulk of the Kuiper belt is confined between Neptune's 3:2 MMR at 40 au and its 2:1 at 48.3 au with moderate eccentricities. 

Our other 11 Kuiper belt formation simulations display dynamical evolution that is broadly similar to that shown in Figure \ref{fig:ExampleSim}. A main difference is the time and Neptunian semimajor axis at which the planetary instability occurs within individual systems. These values range from $\sim$13--58 Myrs and 25.6--29.0 au, respectively, and they are listed in Table \ref{tab:nepmig}. In addition, the details of the rapid evolution of planetary semimajor axes, eccentricities, and inclinations that occurs within the instability are also unique in each simulation. 

\subsection{Pre-Instability Belt Evolution}\label{sec:preinstab}

One notable difference between our work here and other recent models of Kuiper belt formation \citep[e.g.][]{nes15a, kaibshep16, volkmal19} is the starting position of Neptune. These past recent works have mostly studied scenarios in which Neptune migrates from 24--30 au. This initial 24 au position of Neptune immediately places the Neptunian 3:2 MMR at 31.4 au, well beyond the outer edge of the primordial disk at 30 au. In addition, this initial Neptunian orbit also positions the 5:4 and 4:3 MMRs near the outer disk edge at 27.8 and 29.1 au, respectively. Consequently, if Neptune migrates into a planetesimal disk from an initial position of 24 au, much of the material it encounters will never have been swept by a major first order MMR. In contrast, our Figure \ref{fig:ExampleSim} shows that these resonances are major drivers of the early orbital evolution of our primordial belts. These resonances are capable of sweeping up significant fractions of disk material and transporting it as Neptune migrates \citep{mal95, hahnmal05}, and the efficiency of this process is severely curtailed if the resonances never pass over much of the primordial material. It is possible, however, that a lower density disk did extend beyond 30 au, although the lower surface density would imply significantly less material over which the resonances could sweep \citep[e.g.][]{gom04, nes22}.

Meanwhile, our initial conditions are motivated by the success of planetary instability models in which the outer planets' orbits are generated from 5-planet resonant chains \citep{thom99, tsig05, nes11, nesmorb12}, and our chain begins with Neptune orbiting at 20.4 au. (We should note, however, that previous studies of planetary resonant chains did not explicitly consider the resulting Kuiper belt as a metric for compatibility with our solar system.) As mentioned previously, this places the 3:2 MMR initially near the middle of the primordial belt, and the 4:3 and 5:4 lie near the inner edge. Thus, as Neptune moves outward, a much larger portion of the disk is swept by these resonances compared to other recent works modeling Kuiper belt formation. 

\begin{figure}[t]
\centering
\includegraphics[scale=0.4]{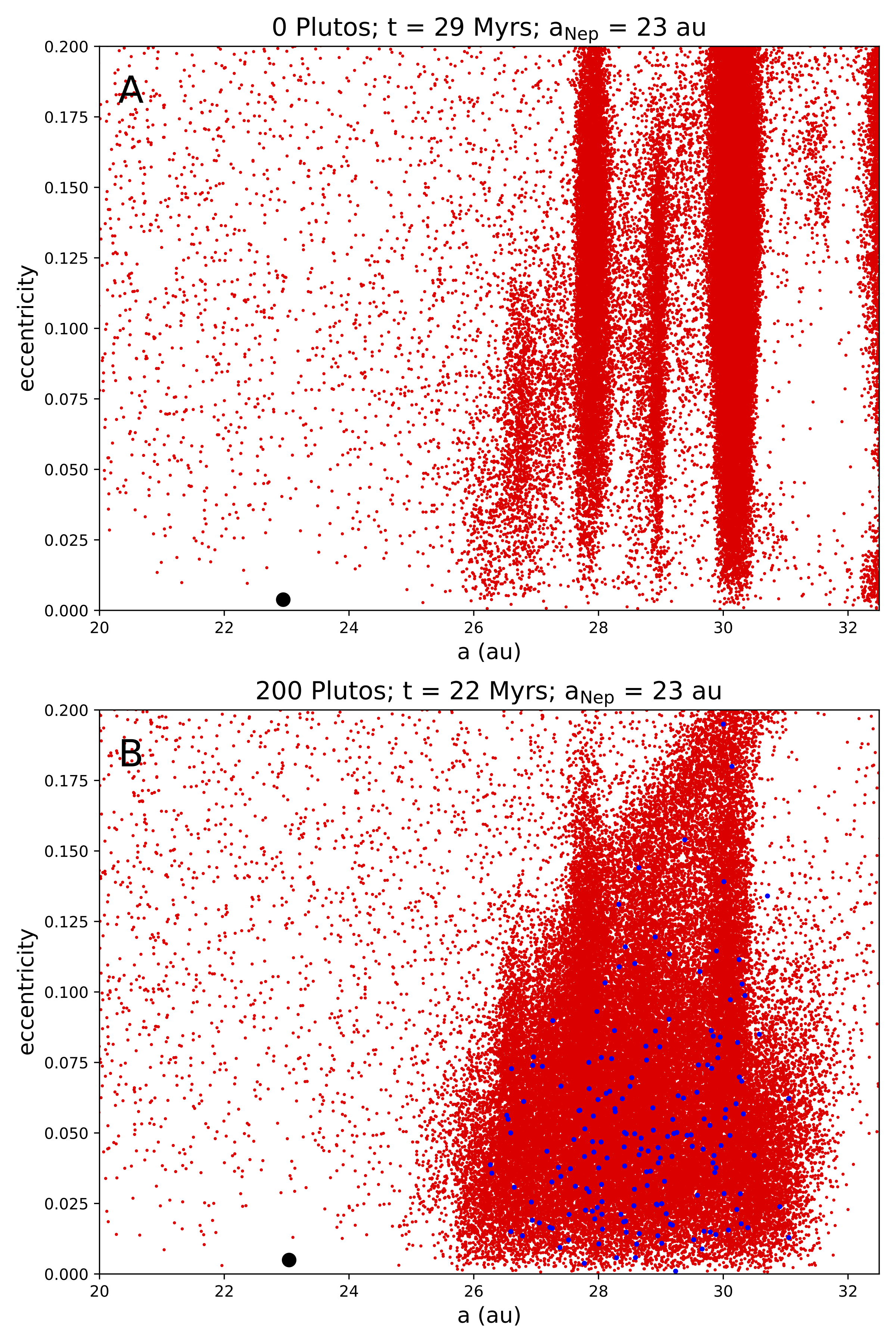}
\caption{Plot of eccentricities vs semimajor axes for all bodies in the 0Pa simulation (Panel A) and the 200Pa simulation (Panel B). Red points mark sub-Plutos, blue points mark Pluto-mass bodies, and the black point marks Neptune.}
\label{fig:selfstir}
\end{figure}

The effects of this resonance sweeping can be seen most clearly in Figure \ref{fig:selfstir}A. Here we plot the eccentricities and semimajor axes of simulation bodies in 0Pa when Neptune has reached 23 au, and its 3:2 MMR has moved to 30.1 au. At this point, the large majority of belt material has been gathered into mean motion resonances with Neptune. 0Pa only contains sub-Pluto bodies, whose interactions between one another are neglected in our simulations. We can see in Figure \ref{fig:selfstir}B how the disk is reshaped when we make an analogous plot for our 200Pa simulation, which contains 200 Pluto-mass bodies that can gravitationally scatter one another as well as the sub-Plutos. In this simulation, we see the same qualitative distribution as 0Pa, but it is now smeared out (this smearing is even more pronounced in other simulations that contain more Pluto-mass bodies). The scattering events between Pluto-mass bodies and other belt members limit long-term storage within resonances and allow more objects to occupy low-eccentricity, inter-resonance orbits. Nonetheless, the sweeping of resonances across the disk reshapes the semimajor axis and eccentricity distributions in both simulations by the time Neptune has reached just 23 au. (For reference, the Kuiper belt simulations of \citet{kaibshep16} assume that Neptune starts at 24 au and migrates into a dynamically cold disk with no sudden changes in semimajor axis distribution.) 

\begin{figure}[t]
\centering
\includegraphics[scale=0.4]{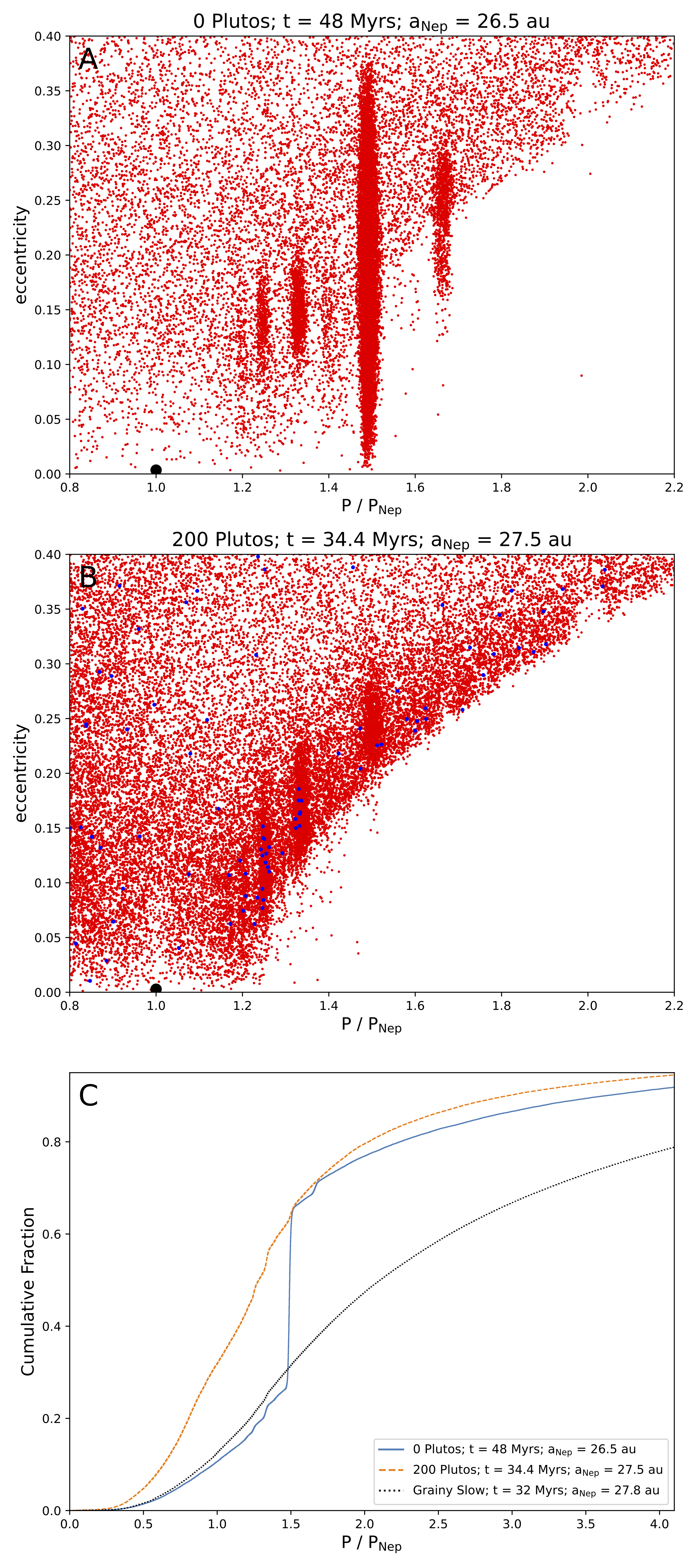}
\caption{{\bf A \& B:} Plot of eccentricities vs period ratio with Neptune for all bodies just prior to the planetary instability in simulation 0Pa (Panel A) and simulation 200Pa (Panel B). Red points mark sub-Plutos, blue points mark Pluto-mass bodies, and the black point marks Neptune. {\bf C:} Cumulative distribution of bodies' period ratio with Neptune just prior to planetary instability and/or Neptunian `jump' in the 0Pa simulation ({\it blue solid}), the 200Pa simulation ({\it orange dashed}), and the grainy slow simulation ({\it black dotted}) from \citet{kaibshep16}.}
\label{fig:selfstir2}
\end{figure}

The effects of this resonance sweeping and gathering persist until the onset of dynamical instability. This can be seen in Figure \ref{fig:selfstir2} where we plot particle eccentricities vs their Neptunian period ratio. (Period ratio is chosen instead of particle semimajor axis because it highlights the positions of resonances and because each simulation's instability occurs at a different Neptunian semimajor axis.) In Figure \ref{fig:selfstir2}A, we see that a very large fraction of particles in 0Pa remain confined to Neptunian mean motion resonances right up until the onset of instability. In Figure \ref{fig:selfstir2}B, we see the concentration within resonances is not as dramatic in the 200Pa simulation, and the regions outside of resonances are more populated with particles. This is because the Pluto-mass bodies prevent most particles from long-term resonance storage. This can also be seen in the absence of low-eccentricity particles in Figure \ref{fig:selfstir2}B's 3:2 resonance, which is positioned at 37 au. The only way for bodies to reach this portion of orbital space is to remain in resonance while the resonance migrates out by 7+ au. Unlike Figure \ref{fig:selfstir2}A, this simply does not happen often, and we might suspect that the resonance sweeping is not an important effect with the inclusion of Pluto-mass bodies. However, when we plot the distribution of particle Neptunian period ratios in Figure \ref{fig:selfstir2}C, we see that most 200Pa particles are still mostly confined Sunward of the 3:2 MMR with Neptune. We may expect this if many particles are temporarily carried within the 3:2 MMR before dropping out at a later time as the resonance continues its march outward. Thus, even though the Neptunian resonances do not gather material as abundantly when Pluto-mass bodies are included, the sweeping of the 3:2 resonance clearly still influences the distribution of particles right up until the planetary instability ensues. The significance of this effect can be better appreciated when we examine the distribution of particle Neptunian period ratios in the ``grainy slow'' run of \citet{kaibshep16} right before Neptune jumps (which is taken to mimic a planetary instability). In this simulation, Figure \ref{fig:selfstir2}C shows that $\sim$1/3 of all particles orbit at or Sunward of Neptune's 3:2 MMR at the time of instability. Meanwhile, in the 0Pa and 200Pa simulations, this number is double that of the \citet{kaibshep16} simulation, or $\sim$2/3 of all particles. 

\begin{figure}[t]
\centering
\includegraphics[scale=0.4]{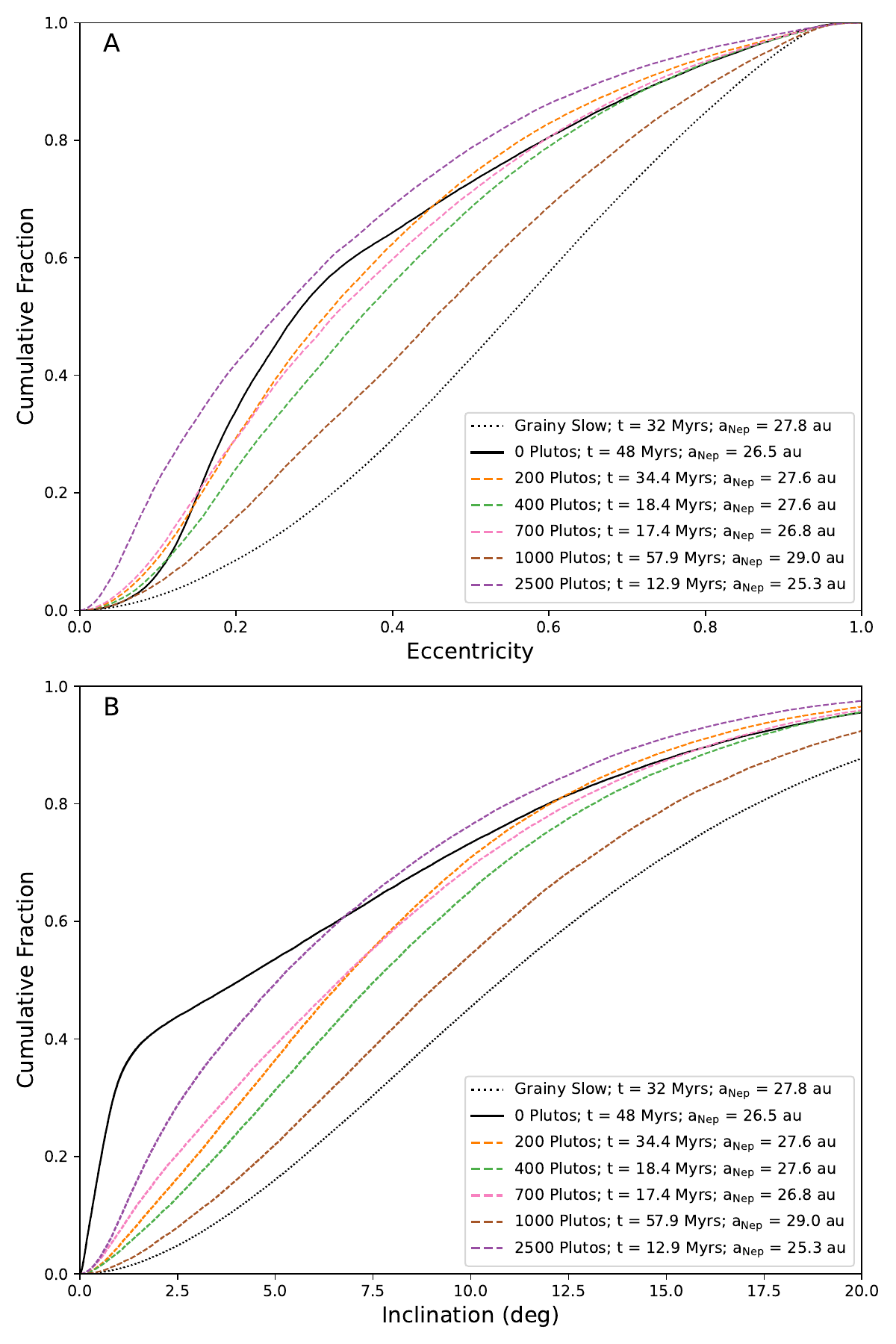}
\caption{Cumulative distribution of particle eccentricities (Panel A) and particle inclinations (Panel B) just prior to planetary instability and/or Neptunian `jump' in the the simulations presented here (see legend) as well as the grainy slow simulation ({\it dotted}) from \citet{kaibshep16}.}
\label{fig:selfstir3}
\end{figure}

The differences between the pre-instability states of the \citet{kaibshep16} ``grainy slow'' run and those of our current set of simulations are not just confined to orbital period ratios. In Figure \ref{fig:selfstir3}A, we plot the distribution of orbital eccentricities just prior to the planetary instability. Here we see that the distribution of eccentricities in the \citet{kaibshep16} grainy slow simulation is systematically hotter than any of the gravitationally driven runs presented in this work. The distributions of our individual gravitationally driven runs are all different because each run contains a unique Neptunian migration and unique numbers of Pluto-mass bodies, but they are all notably more biased toward lower eccentricities than the \citet{kaibshep16} simulation. This is even true of the 1000Pa simulation, which features a Neptunian migration that is larger in both duration and distance than the \citet{kaibshep16} simulation. This difference in eccentricity distribution is again likely due to the effects of the resonance sweeping in our current simulations. While this resonance sweeping can excite particles to moderate eccentricities, it also allows many particles to avoid gravitational scattering encounters with Neptune as Neptune migrates across particles' initial orbital positions in the primordial belt. This means that at the time of instability, we would expect more particles on highly eccentric, large semimajor axis orbits in the \citet{kaibshep16} simulation due to direct scattering events with Neptune, which is what is shown in Figures \ref{fig:selfstir2} and \ref{fig:selfstir3}. 

In Figure \ref{fig:selfstir3}B, we see a similar pattern for particle inclinations. All of our gravitationally driven runs exhibit pre-instability particle inclinations that are lower than the \citet{kaibshep16} grainy slow simulation. This is once again likely a result of the sweeping Neptunian resonances in our current runs allowing particles to avoid Neptune scattering events. The most extreme case is the 0Pa simulation whose efficient resonant storage allows prominent populations of very low-inclination bodies to persist at the 5:4, 4:3, and 3:2 Neptunian MMRs. The median inclination of this simulation is a factor of $\sim$2.5 lower than the \citet{kaibshep16} simulation (4.1$^{\circ}$ vs 10.8$^{\circ}$). The median values of our other runs vary from 5.9--9.8$^{\circ}$, and runs that feature a longer migration distance for Neptune seem to exhibit larger inclinations, but, again, not as large as the \citet{kaibshep16} simulation. Furthermore, the true effect of this resonance shepherding is even stronger than what is implied in Figure \ref{fig:selfstir3}. The reason is that the resonance sweeping saves many more bodies ($\sim$50--60\%) from being ejected prior to instability compared to \citet{kaibshep16}, which only has 35\% of bodies remaining prior to the instability. Thus, in absolute terms, the number of primordial belt objects found on lower eccentricities, inclinations, and semimajor axes before the instability is perhaps a factor of $\sim$1.5 greater than what appears in the cumulative distributions in Figure \ref{fig:selfstir3}.

Thus, if we assume that Neptune forms near or interior to $\sim$20 au as envisioned by many prior works \citep[e.g.][]{nesmorb12, clem18, deien18, nes21b}, low-order mean motion resonances will sweep across nearly all of the primordial planetesimal belt unless a planetary instability occurs before Neptune reaches $\sim$23 au (at which point its 3:2 MMR would sit at $\sim$30 au). Such sweeping is not included in many recent models of the formation of the modern Kuiper belt from a migrating Neptune \citep[e.g.][]{nes15a, kaibshep16, volkmal19}. Consequently, our work here predicts a very different distribution of planetesimal orbits exterior to Neptune when the planetary instability occurs. While the modern Kuiper belt's properties have been used to constrain Neptune's pre-instability behavior, neglecting Neptune's pre-instability resonance sweeping may result in incorrectly mapping Neptune's pre-instability orbital evolution to the resulting modern Kuiper belt properties.

\subsection{Final Kuiper Belt Properties}\label{sec:FinKB}

\begin{table*}
\centering
\begin{tabular}{c}
\textbf{Final Kuiper Belt Properties}\\
\end{tabular}
\\
\begin{tabular}{c c c c c c c c}
\hline
Sim & Trapping \% & HB:Plutino & HB Trapping \% & $\sigma_{\rm HB}$ & N$_{\rm Pluto}$ \\
 & & & & ($^{\circ}$) &\\
\hline
0Pa & 0.43 & $\infty$ & 0.036 & 8.6 & N/A\\
0Pb & 8.12 & $<0.01$ & 0.126 & 15.0 & N/A\\
200Pa & 0.20 & 4.6 & 0.042 & 9.6 & 2\\
200Pb & 0.34 & 1.7 & 0.065 & 9.8 & 1\\
400Pa & 0.52 & 0.52 & 0.068 & 5.6 & 1\\
400Pb & 0.15 & 0.12 & 0.005 & - & 0\\
700Pa & 0.29 & 30 & 0.063 & 4.9 & 1\\
700Pb & 0.39 & 0.89 & 0.063 & 9.7 & 3\\
1000Pa & 0.82 & 23 & 0.124 & 13.5 & 12\\
1000Pb & 0.18 & 0.55 & 0.032 & - & 3\\
2500Pa & 0.52 & 2.8 & 0.123 & 8.7 & 27\\
2500Pb & 0.41 & 11 & 0.074 & 6.8 & 15\\
\hline
\end{tabular}
\caption{Table of properties describing the Kuiper belts formed in our simulated systems. From left to right, the columns are: (1) simulation name, (2) percentage of primordial disk particles surviving at the end of the simulation, (3) ratio of hot Kuiper belt population to Plutino population, (4) percentage of primordial disk particles trapped in the hot Kuiper belt at the end of the simulation, (5) best-fit inclination distribution to the final hot Kuiper belt population (populations under 20 particles are omitted), and (6) number of Pluto-mass bodies surviving at the end of the simulation.}
\label{tab:KBprop}
\end{table*}

\begin{figure}[t]
\centering
\includegraphics[scale=0.4]{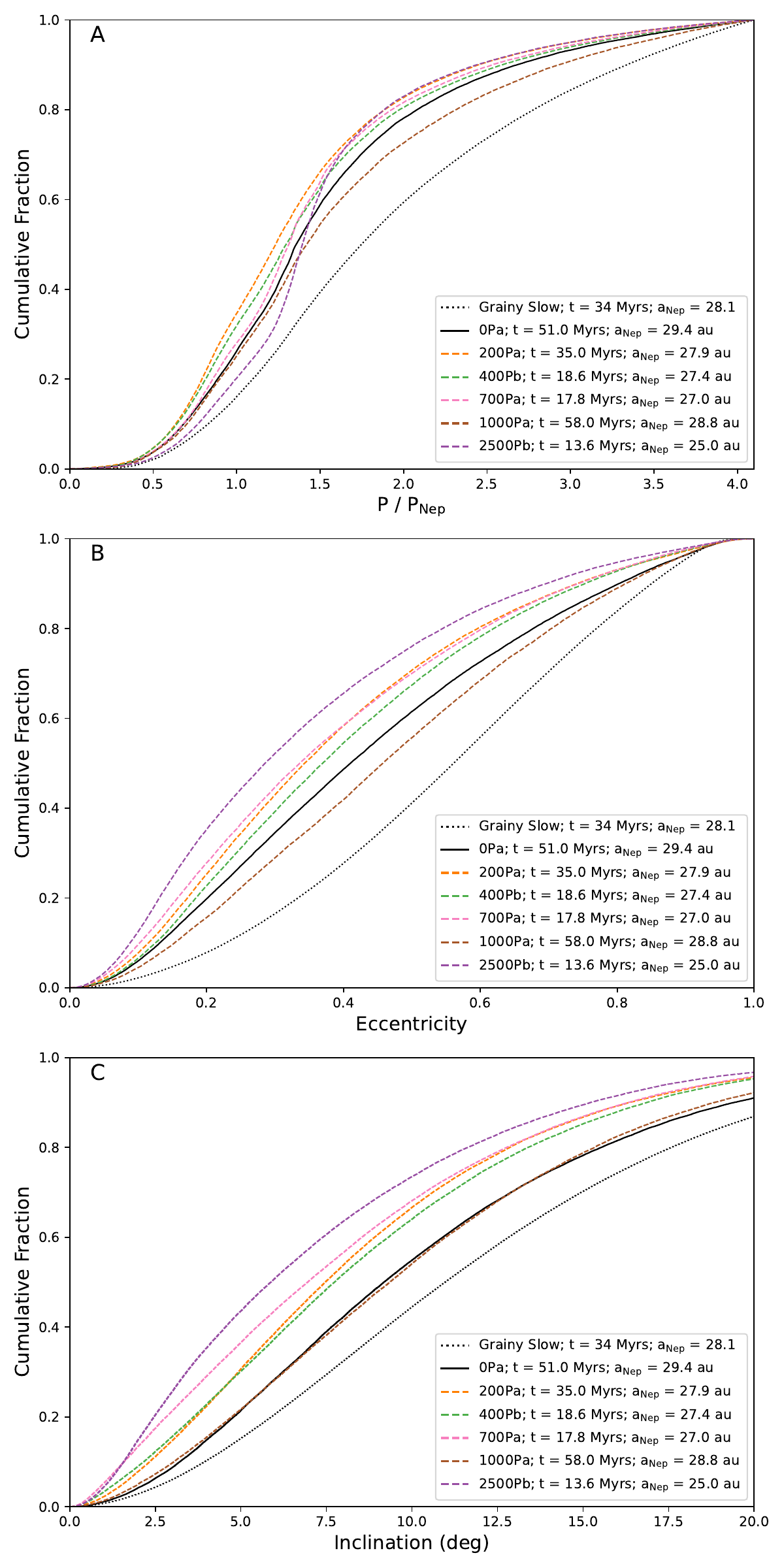}
\caption{Cumulative distribution of particle eccentricities (Panel A) and particle inclinations (Panel B) just after the planetary instability and/or Neptunian `jump' in the the simulations presented here (see legend) as well as the grainy slow simulation ({\it dotted}) from \citet{kaibshep16}.}
\label{fig:postinstab}
\end{figure}

It is now clear that planetesimals in our gravitationally driven systems maintain smaller semimajor axes, eccentricities, and inclinations prior to the planetary instability, but it is possible that the dynamical upheaval of the planetary instability may erase these differences. However, Figure \ref{fig:postinstab} suggests that this is not so. Here we plot the distributions of Neptunian period ratios, eccentricities, and inclinations of selected gravitationally driven simulations $\sim$1 Myrs after their planetary instabilities. These are again compared with the \citet{kaibshep16} grainy slow simulation 1 Myrs after Neptune's jump. We can see that although there is again variance among the gravitationally driven simulations, they all feature a bias toward smaller Neptune period ratios (or semimajor axes), orbital eccentricities, and orbital inclinations than the \citet{kaibshep16} simulation. Thus, the planetesimal distribution differences between our gravitationally driven simulations and the forced migration \citet{kaibshep16} simulation persist after the planetary instability completes and residual planetary migration initiates. 

Previous works have sought to connect the modern orbital architecture of the Kuiper belt with early solar system events and conditions. In particular, the inclination distribution of KBOs as well as the population ratio of Plutinos to the Hot Classical Belt have been argued to constrain Neptune's migration speed as well as its smoothness, which has been taken as a proxy for the number of primordial Pluto-mass bodies \citep{nes15a, nes15b}. Thus, we also examine these Kuiper belt properties in our simulations, and they are listed in Table \ref{tab:KBprop}. 

\subsubsection{Kuiper Belt Inclinations}

One immediate feature of the Kuiper belt properties listed in Table \ref{tab:KBprop} is the large variance in their hot Kuiper belt inclination distributions. We extract particles analogous to Hot Classical KBOs from our simulations by selecting surviving bodies with perihelia at least 6 au beyond Neptune and orbital periods between 1.6--1.9 times Neptune's period. (In the modern solar system, these selection criteria are equivalent to $q>36.1$ au and $41.2 < a< 46.2$ au.) When we fit $\sin i$ times a Gaussian to the inclination distribution of these hot belt particles, we find that the best fit Gaussian's standard deviation ($\sigma_{HB}$) fluctuates by about a factor of 3 across our simulations. $\sigma_{HB}$ can be as low as 5$^{\circ}$, and only two of our simulations (0Pa and 1000Pa) approach the 14$^{\circ}$ value estimated for the real Hot belt \citep{van19}. 

It is not immediately obvious that there is a single driver that primarily determines the final value of $\sigma_{HB}$ in our simulations. There is correlation between the inclination dispersion of Table \ref{tab:KBprop} and Neptune's migration timescale, but it is not particularly strong. Among the timescales listed in Table \ref{tab:nepmig} ($t_{stop}$, $\dot{a}_{25au}$, and $\tau_{post}$), the inclination distribution ($\sigma_{HB}$) displays the highest degree of correlation with $\dot{a}_{25au}$. However, with a Pearson correlation coefficient of $r\simeq0.45$, there are likely other factors at play.

Meanwhile, the primary parameter varied in our initial conditions is the number of primordial Pluto-mass bodies, but there is little correlation ($r\simeq-0.27$) between the number of such bodies and the inclination distribution of the Hot Classical Belt. In some cases, simulations that possess the same initial number of Pluto-mass bodies (700Pa vs 700Pb, or 0Pa vs 0Pb) have $\sigma_{HB}$ values that differ by a factor of $\sim$2 or more from one another. Due to the nature of our pre-instability cloning, systems with the same number of primordial Pluto-mass bodies experience nearly identical pre-instability planetary evolution. Thus, although Figure \ref{fig:selfstir3} shows that this phase of our simulations excites Kuiper belt inclinations, significant additional inclination evolution must take place during and after the instability. 

It is illuminating to specifically examine the detailed evolution of the 700Pa and 700Pb runs. Each simulation traps 59 particles into the Hot Classical Belt, but the final inclinations of these bodies are twice as large in 700Pb ($\sigma_{HB}=9.7^{\circ}$) as in 700Pa ($\sigma_{HB}=4.9^{\circ}$), even though the two simulations have very similar post-instability migration timescales as well. Examining a simulation time output 1 Myr after each system's planetary instability, we find that the bodies destined to become Hot Classical members broadly have orbital perihelia between 20--32 au and semimajor axes between 20--60 au. If we look at all particles in this range of perihelion and semimajor axis at 1 Myr after the planetary instability, we find a median inclination of 6.3$^{\circ}$ in the 700Pa and 8.8$^{\circ}$ in the 700Pb. Meanwhile, just 1 Myr prior to instability these particles' median inclination was 4.2$^{\circ}$. Thus, the median inclination more than doubles during the planetary instability in the 700Pb system, while it only increases by 50\% in the 700Pa system. This strongly hints that the planetary instability itself can be a major source of inclination heating, and we see different behaviors in the two systems' instabilities. In the case of 700Pa, no planet records an inclination above $\sim$$1^{\circ}$ during or after the instability. Meanwhile, in 700Pb's instability, every planet's inclination except Jupiter's exceeds 1$^{\circ}$ at some point, and the lost ice giant attains a $\sim$4$^{\circ}$ inclination prior to ejection, as does Uranus (although dynamical friction eventually drives it back below 1$^{\circ}$). It appears that the excitement of planetary inclinations during the instability can be a significant driver of the ultimate Kuiper belt inclination distribution.

The influence of the giant planet instability can occur through two different mechanisms. First, the same planetary perturbations that are exciting Neptune's orbital inclination can also excite the inclinations of the remaining primordial belt objects. Second, after the instability ceases, Neptune's inclination and eccentricity are damped back down during the planet's residual migration through the Kuiper belt. This requires that the inclinations and eccentricities of the Kuiper belt objects must be further excited as a result. Although many of these will be lost via ejection, some will not and will contribute to the final structure of the Kuiper belt. 

To assess which of these effects may be most important, we rerun the portions of 400Pa and 700Pa immediately after the instability, beginning with the first time output after an ice giant is lost. The original runs of these two simulations yielded the lowest values of $\sigma_{HB}$ (5.6$^\circ$ and 4.9$^\circ$), and they had modest values for Neptune's maximum inclination (0.81$^\circ$ and 0.22$^\circ$). In the post-instability reruns, Neptune is restarted with an inclination of 2$^\circ$ instead, which is roughly the largest maximum recorded for any simulation. When we repeat these runs and examine their states at $t=100$ Myrs, we see very little difference from the original runs. When we extract particles between the 3:2 and 2:1 Neptunian MMR and with perihelia beyond 34 au, we find that a Kolmogorov-Smirnov test cannot reject the null hypothesis ($p$-values of 0.15 and 0.74) that the original and repeated particle inclinations have the same underlying distribution. Thus, this suggests that the planetary dynamical evolution {\it within the instability} is what can drive significant increases in Kuiper belt inclinations in our simulations.

\subsubsection{Primordial Plutos and the Hot Belt to Plutino Ratio}

Table \ref{tab:KBprop} also provides the final ratio of Hot Classical objects to Plutinos in each of our simulations. To be categorized as a Plutino, a surviving simulation body must be located near the semimajor axis of the 3:2 MMR with Neptune and have a librating resonant angle (which was checked by eye for all candidate particles, except for 0Pb, which has thousands of bodies near the 3:2). Like $\sigma_{HB}$, the ratio of Hot Classical bodies to Plutinos varies wildly from run to run. In three of our simulations, it exceeds 20:1, while in another it is less than 1:100. Previously, it has been suggested that this ratio is dependent on the number of Pluto-mass objects in the primordial Kuiper belt, as the graininess of Neptune's migration induced by encounters with Pluto-mass objects can drive resonant dropout into the Hot Classical Belt \citep{zhou02, nes16}. For a single, assumed giant planet orbital evolution sequence, this dependency is likely significant. However, each of our runs has a unique temporal evolution of the giant planets' orbits. When we examine the ratio of Hot Classical bodies to Plutinos in our simulations, there is no obvious correlation with the number of Pluto-mass bodies we employ. In fact, pairs of runs with the same initial numbers of Pluto-mass bodies (0, 700, and 1000) can have final ratios that differ by well over an order of magnitude. 

Clearly, there must be other factors at play in determining the ratio of Hot Classical belt objects to Plutinos. An obvious one is the behavior of Neptune during the planetary instability. We can see this when we examine our two runs without Plutos (0Pa and 0Pb). These two runs actually return our highest and lowest Hot Belt-to-Plutino ratios. 0Pb finishes with $\sim$7-8\% of all initial objects locked in the 3:2 resonance, comprising over 90\% of the surviving Kuiper belt. This is because 0Pb features no planetary encounters with Neptune large enough to induce a significant change in semimajor axis or eccentricity during the planetary instability. As a result, virtually all of the particles swept and stored into the 3:2 MMR during Neptune's pre-instability migration can remain there after the instability. In contrast, in 0Pa, Neptune encounters the ejected ice giant during the instability, which causes it to jump 2.2 au out to 29.2 au and attain an eccentricity of 0.12. Thus, any resonant population is lost and capture must start anew at 0 near the outer edge of the original primordial disk. While other simulations feature rebuilt Plutino populations after sudden changes to Neptune's orbit (see 400Pa and 700Pb), the stability of Pluto-like orbits over the age of the solar system display subtle dependencies on giant planet architecture \citep{malito22}, and none of our final architectures are identical. 

As 0Pb illustrates, the number of Plutinos trapped into the modern Kuiper belt has the potential to be huge if Neptune's orbit is not perturbed. However, this simulation likely overstates the case, as it only contains semi-active sub-Pluto bodies. Simulation 400Pb also features an instability that does not significantly alter Neptune's semimajor axis. Here, however, the final population of Plutinos falls by a factor of $\sim$100 (even though it is still 10x greater than the Hot Belt population). The reason for this is that the presence of Pluto-mass bodies limit the long-term storage of bodies in Neptunian resonance. This can be seen in Figure \ref{fig:selfstir2}C where we compare the pre-instability semimajor axis distribution of planetesimals in 200Pb and 0Pb. It is clear that the semimajor axis distribution is qualitatively similar but smeared in runs with significant numbers of Pluto-mass bodies. 

Pluto-mass bodies may be able to erode the population of bodies trapped in Neptunian resonance in two ways. As discussed previously, the first is through the ``jitter'' they impart on Neptune's semimajor axis evolution when they undergo close encounters with the planet. This jitter causes sudden changes to the location of Neptune's resonances, which can lead to resonant dropout, and this mechanism has been studied before \citep{zhou02, nes16}. The second mechanism occurs during direct encounters between Pluto-mass bodies and the objects occupying resonances. Just as Pluto-mass encounters generate semimajor axis changes for Neptune, we expect they will also do this for Kuiper belt objects \citep{hahnmal05}. 

To highlight this second resonance removal mechanism and compare it with the first, we turn to the well-behaved pre-instability epoch of one of our 0Pa/b simulations. In the original run, over the course of $\sim$60 Myrs, Neptune migrates from $\sim$20 au to $\sim$27 au interior to a disk of 10$^5$ sub-Pluto bodies that do not interact with one another. Near the end of this migration phase, nearly 40\% of disk bodies have become swept up near Neptune's 3:2 resonance. To study how Pluto-mass bodies can alter the resonance trapping during this phase, we repeat this phase of the simulation two more times with 200 randomly selected sub-Pluto bodies promoted to Pluto-mass bodies. In the first repeat run, Pluto-mass bodies are treated as semi-active and therefore still do not interact with the sub-Plutos (or one another), but they do interact with Neptune, inducing a migration jitter. In the second repeat, the Pluto-mass bodies are fully active, allowing them to gravitationally interact with all bodies.

\begin{figure*}[t]
\centering
\includegraphics[scale=0.6]{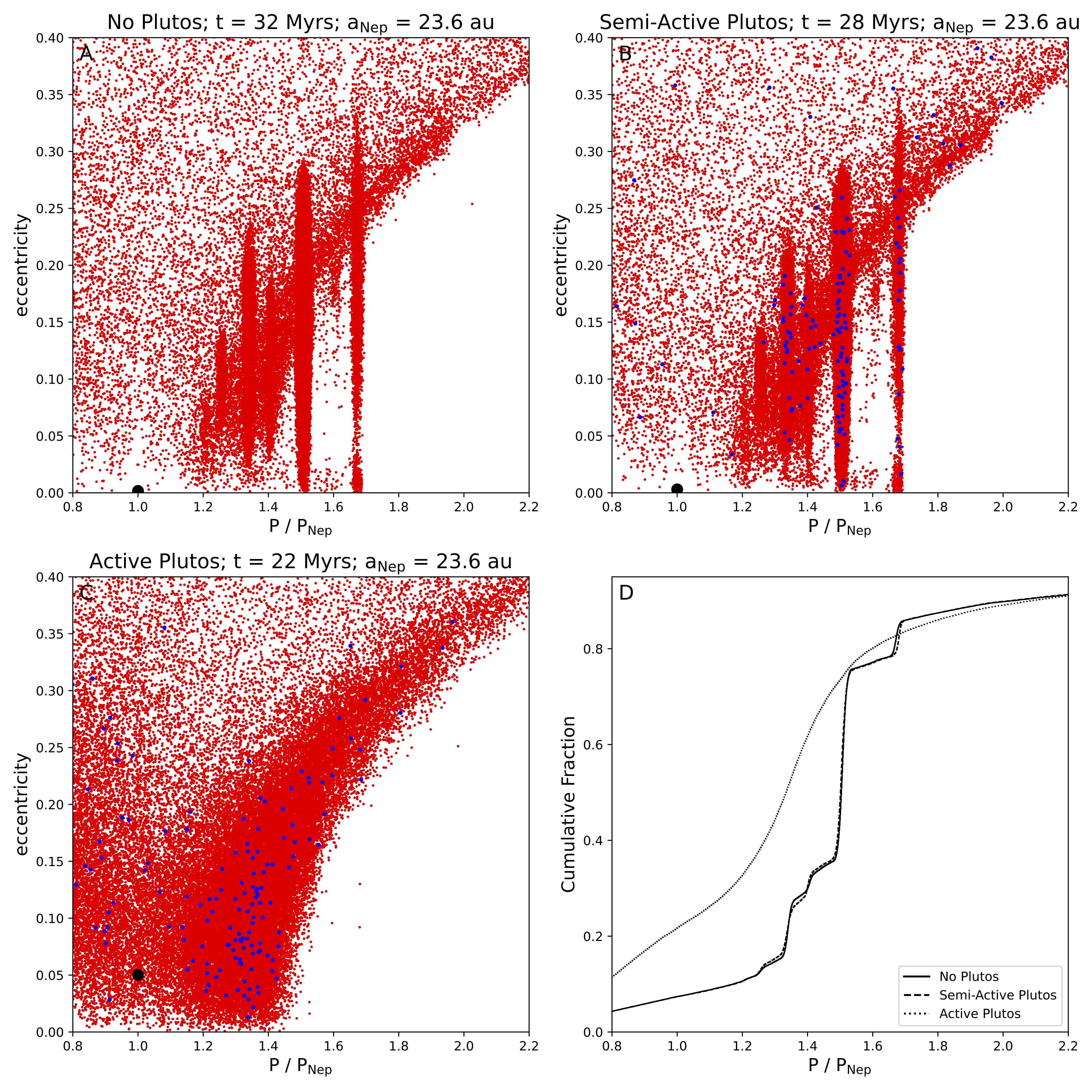}
\caption{{\bf A--C:} Particle eccentricities vs orbital period ratio with Neptune for the 0Pa/b simulation ({\it panel A}) as well as a repeated version with 200 semi-active Pluto-mass bodies ({\it panel B}) and another repeated version with 200 fully active Pluto-mass bodies ({\it panel C}). Sub-pluto particles with masses of 0.09 M$_{Pluto}$ are marked with red points, Pluto-mass particles are marked with blue points, and Neptune is marked with a black point. Each simulation snapshot is taken when Neptune's semimajor axis reaches 23.6 au, and the simulation times are marked above each panel. {\bf D:} The cumulative distribution of particles' orbital period ratios with Neptune is plotted for our 0Pa/b simulation ({\it solid}) as well as its repeat with 200 semi-active Pluto-mass bodies ({\it dashed}) and its repeat with 200 fully active Pluto-mass bodies ({\it dotted}).}
\label{fig:PlutoEffects_preinstab}
\end{figure*}

Snapshots of the original simulation as well as its two repeats are shown in Figure \ref{fig:PlutoEffects_preinstab} when Neptune has reached 23.6 au (after which the system with fully active Plutos passes through an instability). As can be seen in panels A and B, the original simulation and the simulation with 200 semi-active Plutos look very similar by eye. However, there is some evidence that the presence of Pluto-induced migration jitter has resulted in modest resonance dropout. The original 0Pa/b run has 31.1 percent of its bodies with orbital periods between 1.47--1.53 of Neptune's period. This percentage falls to 29.2\% in the repeat with 200 semi-active Plutos. However, in the repeat with 200 fully active Plutos the differences are dramatic in the snapshots, and the percentage of bodies with orbital periods between 1.47--1.53 of Neptune's period falls by an order of magnitude to 3.7\%. Thus, it appears that direct scattering between Pluto-mass bodies and other primordial belt bodies are much more effective at removing objects from resonances, at least prior to the planetary instability.

A lucky occurrence in our first repeat of the 0P simulations also allows us to study the mechanisms enhancing resonance dropout during the post-instability regime as well. In this simulation, the planetary instability caused a collision between the innermost ice giant and Saturn, and the orbits of Uranus and Neptune were not disturbed. Although Neptune and Uranus migrate well beyond their 2:1 resonance (limiting this simulation's relevance to our real planets' migration), a large population of Plutinos is maintained during Neptune's slower, residual migration. This system retains 90 of the 200 particles we manually graduated to Pluto-masses. We run the post-instability, post-2:1-crossing phase of this simulation three times: once neglecting the interactions between Pluto-mass bodies and other small bodies, another time including these interactions, and a final time with all of our Pluto-mass bodies again demoted to semi-active sub-Plutos. 

\begin{figure*}
\centering
\includegraphics[scale=0.6]{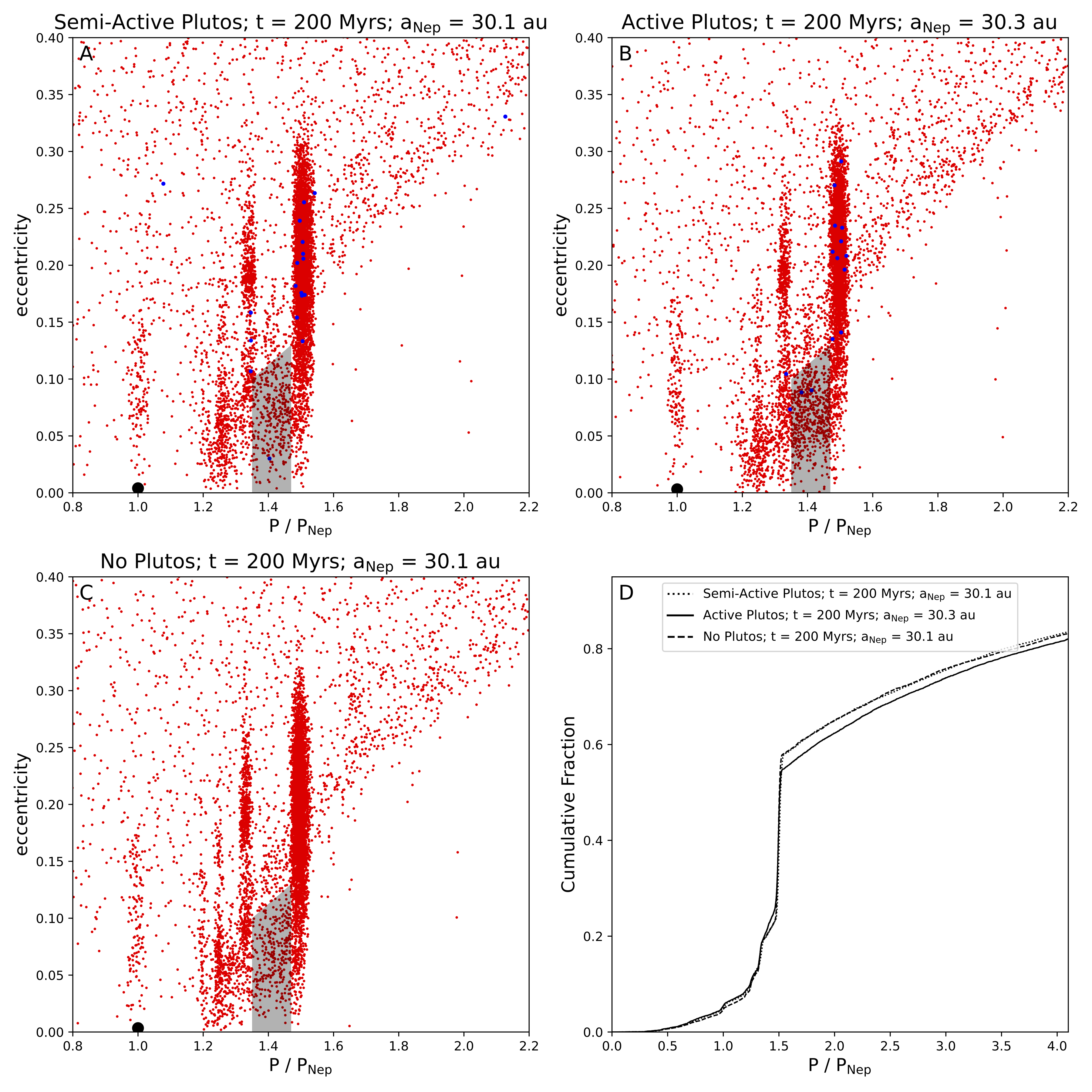}
\caption{{\bf A--C:} Particle eccentricities vs orbital period ratio with Neptune after 200 Myrs for a simulation with 200 semi-active Pluto-mass bodies ({\it panel A}), another repeated version with 200 fully active Pluto-mass bodies ({\it panel B}), and a final repeat with only sub-Pluto bodies ({\it panel C}). Sub-pluto particles with masses of 0.09 M$_{Pluto}$ are marked with red points, Pluto-mass particles are marked with blue points, and Neptune is marked with a black point. Neptune's semimajor axis is marked above each panel. {\bf D:} The cumulative distribution of particles' orbital period ratios with Neptune is plotted for our simulation with 200 semi-active Pluto-mass bodies ({\it dotted}) as well as its repeat with 200 fully active Pluto-mass bodies ({\it solid}) and its repeat with no Pluto-mass bodies ({\it dashed}).}
\label{fig:PlutoEffects}
\end{figure*}

The results of these three runs at $t=200$ Myrs are shown in Figure \ref{fig:PlutoEffects}. In each system, Neptune has largely completed its outward migration. In panels A--C, we see that the broad Kuiper belt morphology looks similar for all three systems. However, there are subtle differences. We highlight a box of orbital space below $e\sim0.13$ between the 4:3 and 3:2 Neptunian resonances. This region is noticeably more populated with bodies in the run where Pluto-mass bodies are treated as fully active compared to the semi-active Pluto run and the run without any Plutos. The reason for this is that more bodies fall out of the 3:2 resonance when the gravitational effects of Pluto-mass bodies are fully included. 

We can see this when we examine the population of bodies near the 3:2 resonance after 200 Myrs (see Figure \ref{fig:PlutoEffects}D). In the repeat with no Pluto-mass bodies included, there are 4499 bodies with perihelion beyond 25 au and a Neptunian period ratio between 1.47--1.53. When we include the Pluto-mass bodies as semi-active bodies that can only perturb the planets' orbits, this population falls slightly by $\sim$6\% to 4245. This is consistent an increased resonance dropout due to a graininess to Neptune's migration. Finally, in the repeat with fully active Pluto-mass bodies, the population of this region of orbital space is over 20\% lower, at 3583 bodies. This suggests that including the perturbations that Pluto-mass bodies exert on other Kuiper belt bodies will increase resonance dropout by perhaps 3--4 times compared to simulations that only consider the influence Pluto-mass bodies have on the giant planets. 

Since resonant dropout is significantly enhanced when the full perturbations of Pluto-mass bodies are considered, previous estimates of the primordial population of Pluto-mass bodies inferred from resonant dropout of Plutinos may need to be revised downward. Indeed, \citet{nes16} estimates a primordial population of 1000--4000 Pluto-mass bodies, but this only considered resonant dropout from Neptune's migration jitter and did not account for the ability of Pluto-mass bodies to directly dislodge objects from resonance. It is clear from our 2500Pa/b simulations that 1000--4000 Pluto-mass bodies would lead to far too many Pluto-mass bodies in the modern Kuiper belt. Pluto and Eris are likely the only two $\sim$Pluto-mass bodies within 100 au of the Sun \citep{brown15}, while the final states of our 2500Pa and 2500Pb simulations respectively yield 15 and 27 such bodies. While our 1000Pb simulation yields 3 Pluto-mass bodies after 4 Gyrs of evolution, our 1000Pa simulation yields 12. Meanwhile, our 200Pa and 200Pb simulations yield 2 and 1 Pluto-mass bodies, respectively. It should be noted that the 200Pa/b simulations also yield Hot belt:Plutino population ratios ($\sim$2--4) that are also in rough agreement with observations \citep{nes16}. Thus, it appears that the primordial Kuiper belt could have had as few as $\sim$200 Pluto-mass bodies and no more than $\sim$1000. 

\subsubsection{Other Kuiper Belt Subpopulations}

Even with $\sim$10$^5$ initial particles in our simulations, the analysis of our other final Kuiper belt subpopulations is statistically limited. However, some other notable trends do emerge. First, we consider the population of scattering objects. To do this, we compare planetesimals' final semimajor axes with their semimajor axes 10 Myrs prior to the ends of the simulations to search for semimajor axis changes over 1.5 au. In these instances, planetesimals are classified as scattering \citep{shank13}. When we compare the population ratio of the Hot belt to the scattering bodies, we find that the ratio varies between 0.4 to 3.7 across our simulation ensemble, and 75\% of our simulations have a ratio over 1. This contrasts with observational estimates for the real solar system, which put the ratio near 0.4, the very lower end of our simulation range \citep{pet11, law18}. 


We also study variations in the population of Neptunian Trojans in our simulations. Trojan candidates must have final semimajor axes within 5\% of Neptune, and once candidates are identified, their resonant angles are checked for libration by eye. We find a large variance in the rate that Trojans are captured in our simulations, even though they are always smaller than the Hot belt and Plutino populations in every simulation. Five of our simulations possess no Trojans at the end, and another three systems contain just 1--2. In contrast, three systems (0Pb, 400Pa, and 2500Pa) have 20+ Trojans, so the Trojan capture efficiency can vary by over an order of magnitude. It is not clear what accounts for this large variance.

Finally, we consider cold classical belt objects. The cold classical Kuiper belt is thought to be the only Kuiper belt subpopulation that formed in-situ (residing between roughly 43--48 au). Given its modern low inclinations and eccentricities, its non-excitation also serves as a useful constraint on Kuiper belt formation models. However, the initial disks of our simulations are truncated at 30 au, and they therefore do not include an in-situ cold classical Kuiper belt. Nonetheless, we can consult past works studying the connection between the giant planet instability and the excitation of the cold classical belt to infer whether our simulated systems would over-excite the cold belt \citep{bat11, dawclay12, bat12, wolff12}, In particular, it appears that Neptune's eccentricity must stay below 0.12-0.15 to avoid overexciting the cold classical belt's eccentricities \citep{dawclay12, wolff12}. Reviewing Table \ref{tab:nepmig}, we see that only one system (0Pa) records a maximum Neptune eccentricity of 0.12, and the rest remain below this value, so it appears that Neptune generally will not overexcite the eccentricities of the cold classical belt. 

However, we also note previously that the ejected ice giant appears to significantly contribute to the excitation of our Kuiper belt inclinations in some instances. This may have important ramifications for the cold classical belt as well, since one of its defining features is a systematically colder inclination distribution than the rest of the Kuiper belt. \citet{bat12} noted that ejected ice giants can often overexcite the inclinations of cold classical belt objects. Thus, the observed low inclinations of the cold classical belt may exclude instances in our simulations when the ejected ice giant is a significant contributor to the rest of the Kuiper belt's excited inclination distribution. This particular aspect of the planetary instability will be studied in more detail in a future work.

\section{Summary and Conclusions}\label{sec:con}

In this work, we perform 12 simulations that model the formation of the Kuiper belt from an era shortly after the dissipation of the gaseous component of the Sun's protoplanetary disk until the modern epoch. Our systems initiate Jupiter, Saturn, and three ice giants in a 3:2, 3:2, 2:1, 3:2 resonant configuration surrounded by a 20 M$_{\oplus}$ primordial belt of $\sim$10$^5$ planetesimals extending out to 30 au. These simulations are then evolved for 4 Gyrs, during which the planets unlock from resonance, migrate until the onset of an orbital instability, and complete a post-instability phase of residual orbital migration. This planetary orbital evolution greatly depletes the planetesimal population, but a small fraction (typically 10$^{-3}$--10$^{-2}$) survives in a reservoir reminiscent of our solar system's Kuiper belt. 

Compared to most other 5-planet configurations, the planetary resonant configuration we choose for our simulations is known to display an elevated level of success in replicating the modern orbits of the giant planets after passing through instability \citep{nesmorb12}. The final planetary orbits generated from our orbital instabilities are broadly similar to past instability studies, in that $e_{55}$ is under-excited and $e_{66}$ is over-excited. In addition, we find that Uranus and Neptune are very likely to migrate across their 2:1 MMR (11 of 12 runs), which necessitated us manually moving the location of Uranus after the instability. (This unexpected feature may warrant further study in future work.) Finally, we find that Neptune's semimajor axis migration is not well-modeled with an exponential decay with a fixed timescale. Pre-instability, the migration curve is nearly linear and sometimes even convex. Post-instability, an exponential fit underestimates the migration rate initially and overestimates the rate at late times. Neptune still has to migrate its final 0.1 au after a median time of 340 Myrs in our simulations. 

Because we model the formation of the Kuiper belt starting from a favored 5-planet resonant configuration, Neptune's 5:4, 4:3, and 3:2 MMRs all sweep across much of the primordial belt over the first few au of Neptune's migration. This causes many of the planetesimals to be shepherded outward as Neptune begins its outward migration, and they avoid direct encounters with Neptune. As a result, at the onset of the orbital instability, more of our planetesimals are surviving, and more are found on orbits with smaller semimajor axes, eccentricities, and inclinations compared to recent works that model the formation of the Kuiper belt with a more distant starting orbit for Neptune. These differences persist through the epoch immediately following the planetary instability and may help explain why the properties of the Kuiper belts formed here differ significantly from other recent works. 

In addition, the final Kuiper belts that our simulated systems generate display a large amount of stochasticity. This appears to be at least in part connected with the highly chaotic dynamical evolution that takes place as the planets' orbits evolve within the instability. We see evidence that instabilities that temporarily excite the inclinations of the planets can also significantly excite the inclination distribution of the Kuiper belt, although nearly all of our simulations finish with lower inclinations than the observed belt. In addition, Neptune's semimajor axis often jumps during the instability (both Sunward and anti-Sunward), and this jump leads to losses of resonant bodies, influencing the relative fraction of the final Kuiper belt population trapped in resonance. 

Finally, the primordial belts of our simulations begin with varying numbers of Pluto-mass bodies, from 0--2500, and these bodies have a significant impact on the dynamics of the belt. In their absence, a huge fraction of primordial belt objects become trapped in Neptune resonances before the planetary instability. However, just the modest addition of 200 Pluto-mass bodies lowers this fraction by over an order of magnitude. Moreover, the most powerful way that Pluto-mass bodies remove objects from resonance is through directly perturbing them out of the resonance via gravitational encounters, a mechanism that is absent from many past simulations of Kuiper belt formation. Resonance loss resulting from small changes of Neptune's semimajor axis during encounters with Pluto-mass bodies also occurs, but at a rate that is 3--4 times lower. In addition, 3 of our 4 simulations that begin with 1000 or more Pluto-mass bodies finish with 12 or more Pluto-mass bodies in their final Kuiper belts, while our two simulations with 200 Pluto-mass bodies finish with 1 and 2 Pluto-mass bodies. This suggests that our real Kuiper belt's population of two close-in Pluto-mass bodies (Pluto \& Eris) is consistent with a primordial population of Pluto-mass bodies of no more than $\sim$1000 and perhaps as few as $\sim$200 or less. 

\section{Acknowledgements}

We thank the two anonymous reviewers for their constructive feedback on our work. NAK's contributions were supported from NSF CAREER award 2405121 and NASA Emerging Worlds grant 80NSSC23K0771. MSC is supported by NASA Emerging Worlds grant 80NSSC23K0868 and NASA's CHAMPs team, supported by NASA under Grant No. 80NSSC21K0905 issued through the Interdisciplinary Consortia for Astrobiology Research (ICAR) program. This work used the Expanse GPU cluster at the San Diego Computing Center through allocation AST190052 from the Advanced Cyberinfrastructure Coordination Ecosystem: Services \& Support (ACCESS) program, which is supported by National Science Foundation grants \#2138259, \#2138286, \#2138307, \#2137603, and \#2138296. In addition, this work also made use of the High Performance Computing Resource in the Core Facility for Advanced Research Computing at Case Western Reserve University as well as the OU Supercomputing Center for Education \& Research (OSCER) at the University of Oklahoma (OU).\clearpage

\bibliographystyle{apj}
\bibliography{KBPlutos}

\end{document}